\documentclass{aa}
\usepackage{txfonts}
\usepackage{graphicx}
\usepackage{booktabs}
\usepackage{amsmath}
\usepackage{colortbl}
\usepackage{gensymb}
\usepackage{subfig}
\usepackage{xcolor}
\usepackage{rotating}
\usepackage{amssymb}
\usepackage{latexsym}
\usepackage{ifthen}
\usepackage{enumitem}

\def\lsimeq{\hbox{\raise0.5ex\hbox{$<\lower1.06ex\hbox{$\kern-1.07em{\sim}$}$}}} 
\def\gsimeq{\hbox{\raise0.5ex\hbox{$>\lower1.06ex\hbox{$\kern-1.07em{\sim}$}$}}} 

\begin{document}

   \title{Molecular outflow and feedback in the obscured quasar XID2028 revealed by ALMA}

   \author{
M. Brusa \inst{1,2} 
\and 
G. Cresci\inst{3}
\and E. Daddi \inst{4}
\and R. Paladino\inst{5}
\and M. Perna\inst{3}
\and A. Bongiorno\inst{6}
\and E. Lusso\inst{7}
\and  M. T. Sargent\inst{8}
\and  V. Casasola\inst{3}
\and  C. Feruglio\inst{9}
\and  F. Fraternali\inst{1}
\and  I. Georgiev\inst{10}
\and V. Mainieri\inst{11}
\and  S. Carniani\inst{12,13}
\and  A. Comastri\inst{2}
\and F. Duras\inst{6}
\and F. Fiore\inst{6}
\and F. Mannucci\inst{3}
\and  A. Marconi\inst{14,3}
\and E. Piconcelli\inst{6}
\and  G. Zamorani\inst{2}
\and R. Gilli\inst{2}
\and  F. La Franca\inst{15}
\and  G. Lanzuisi\inst{1,2}
\and D. Lutz\inst{16}
\and  P. Santini\inst{6}
\and N. Z. Scoville\inst{17} 
\and  C. Vignali\inst{1,2}
\and  F. Vito\inst{18,19}
\and  S. Rabien\inst{16}
\and  L. Busoni\inst{3}
\and  M. Bonaglia\inst{3}
}

\institute{Dipartimento di Fisica e Astronomia, Universit\`a di Bologna,  via Gobetti 93/2,  40129 Bologna, Italy 
\and INAF - Osservatorio Astronomico di Bologna, via Gobetti 93/3,  40129 Bologna, Italy 
\and INAF - Osservatorio Astrofisico di Arcetri, Largo Enrico Fermi 5, 50125 Firenze, Italy 
\and CEA, IRFU, DAp, AIM, Universit\'e Paris-Saclay, Universit\'e Paris Diderot,  Sorbonne Paris Cit\'e, CNRS, F-91191 Gif-sur-Yvette, France
\and INAF - Istituto di Radioastronomia, via Gobetti 101, 40129 Bologna, Italy
\and INAF - Osservatorio Astronomico di Roma, via Frascati 33,   00078 Monte Porzio Catone (RM) Italy 
\and Centre for Extragalactic Astronomy, Department of Physics, Durham University, South Road, Durham, DH1 3LE, UK
\and Astronomy Centre, Department of Physics and Astronomy, University of Sussex, Brighton, BN1 9QH, UK
\and INAF - Osservatorio Astronomico di Trieste, via G.B. Tiepolo, 11, I-34143 Trieste, Italy
\and Max Planck Institut f\"ur Astronomie, K\"onigstuhl 17, Heidelberg 69117, Germany
\and  European Southern Observatory, Karl-Schwarzschild-str. 2,  85748 Garching bei M\"unchen, Germany 
\and Cavendish Laboratory, University of Cambridge, 19 J. J. Thomson Ave., Cambridge CB3 0HE, UK
\and Kavli Institute for Cosmology, University of Cambridge, Madingley Road, Cambridge CB3 0HA, UK
\and Dipartimento di Fisica e Astronomia, Universit\`a di Firenze, via G. Sansone 1, I-50019 Sesto F.no (Firenze), Italy
\and Dipartimento di Matematica e Fisica, Universit\`a Roma Tre, via della Vasca Navale 84, I-00146 Roma, Italy 
\and Max Planck Institut f\"ur Extraterrestrische Physik, Giessenbachstrasse 1, 85748 Garching bei M\"unchen, Germany 
\and California  Institute  of  Technology,  MC  249-17,  1200  East California Boulevard, Pasadena, CA 91125, USA
\and Department of Astronomy \& Astrophysics, 525 Davey Lab, The Pennsylvania State University, University Park, PA 16802, USA
\and Institute for Gravitation and the Cosmos, The Pennsylvania State University, University Park, PA 16802, USA
     }
   \date{Received 26 July 2017 ; accepted 12 December 2017}

\abstract
{We imaged, with ALMA and ARGOS/LUCI, the molecular gas and dust and stellar continuum in XID2028, which is an obscured quasi-stellar object (QSO) at z=1.593, where the presence of a massive outflow in the ionised gas component traced by the [O III]5007 emission has been resolved up to 10 kpc. This target  represents a unique test case to study QSO feedback in action at the peak epoch of AGN-galaxy co-evolution. The QSO was detected in the CO(5-4) transition and in the 1.3 mm continuum at $\sim$30 and $\sim$20$\sigma$ significance, respectively; both emissions are confined in the central ($<4$ kpc) radius area.  
Our analysis suggests the presence of a fast rotating molecular disc (v$\sim 400$ km s$^{-1}$) on very compact scales well inside the galaxy extent seen in the rest-frame optical light ($\sim10$ kpc, as inferred from the LUCI data).  
Adding available measurements in additional two CO transitions, CO(2-1) and CO(3-2), we could derive a total gas mass of $\sim10^{10}$ M$_\odot$, thanks to a critical assessment of CO excitation and the comparison with the Rayleigh-Jeans continuum estimate. This translates into a very low gas fraction ($<5$\%) and depletion timescales of 40-75 Myr,  reinforcing the result of atypical gas consumption conditions in XID2028, possibly because of feedback effects on the host galaxy. 
Finally, we also detect the presence of high velocity CO gas  at $\sim5\sigma$, which we interpret as a signature of galaxy-scale molecular outflow that is spatially coincident with the ionised gas outflow. XID2028   therefore represents a unique case in which the  measurement of total outflowing mass, of $\sim500-800$ M$_\odot$ yr$^{-1}$ including the molecular and atomic components in both the ionised and neutral phases, was attempted for a high- z QSO. 
}

\titlerunning{ALMA observations of XID2028}
   \maketitle

%

\section{Introduction}

It is now well established that probably all massive galaxies host a supermassive black hole (SMBH; M=10$^{6}-10^{9}$ M$_\odot$) at their  centre and that the mass of these dark objects correlates well with the properties of the host galaxies (see \citealt{Kormendy2013}, and references therein). This implies that some mechanism had to link the small central regions, where the gravitational field of the SMBH dominates, to the larger scales in which the influence of the central objects is expected to be negligible. 

Theoretical models predict that the energy deposited via shocks by accretion disc winds propagates into the galaxy interstellar medium (ISM) during a feedback phase (e.g. \citealt{King2010,Fabian2012,FG2012,Costa2015}). This phase should be characterised by fast winds and should have a strong impact on the gas reservoir of the host galaxy (se e.g. \citealt{DiMatteo2005,Hopkins2008}). Indeed, given the large scales and velocities expected (up to a few 1000 km s$^{-1}$) and the corresponding associated kinetic energies (10$^{43-45}$ ergs s$^{-1}$),  massive galaxy-scale outflows can easily strip and/or heat gas from/in galaxies at a rate comparable to, or larger than, the rate at which the stars are forming in the galaxy; this could potentially solve the long-standing question of how star formation in massive galaxies is quenched (e.g. \citealt{Croton2006}). 

Outflow phenomena can be observationally probed by measuring velocity shifts of absorption or emission lines with respect to the rest-frame velocity, which cannot be simply related to ordered motions in the galaxy. On galactic scales, spectral features tracing the kinematics of large scale energetic winds with velocities as large as 1000-2000 km s$^{-1}$ are now routinely observed in various gas phases (neutral and ionised, atomic and molecular) in the optical, near, and far-infrared, and millimetre bands (e.g. \citealt{Feruglio2010,Sturm2011,Maiolino2012,Weiss2012,Harrison2012,Rupke2013,Cicone2014,Genzel2014,Harrison2014,Brusa2015,Cicone2015,Emonts2016, Zakamska2016_Xshooter,Nesvadba2016,Carniani2015,Bischetti2017,Rudie2017,Veilleux2017,Talia2017}). Moreover, there is now growing evidence that these powerful outflows may affect their host galaxies and suppress star formation in regions impacted by outflows (`negative feedback'; e.g. \citealt{CanoDiaz2012,Cresci2015,Carniani2016,Vayner2017}).

The relation between the outflow episodes seen in various gas phases or tracers is now being investigated over an increasingly large number of sources. 
For example, in Mrk 231, which is a composite QSO-Starburst system at z=0.04217 and a well-studied example of an object with AGN-driven outflows, there are at least two molecular outflow components with velocities of $\sim$800 km s$^{-1}$ at sub-kpc scales \citep{Feruglio2015}. Similarly, using ALMA\ \citet{GarciaB2014} detected AGN-driven outflows in five different molecular gas tracers in the most studied
low-luminosity AGN NGC 1068 (z = 0.00379); these molecular gas tracers have  $\sim$80 km s$^{-1}$ velocities at the spatial resolution of tens of pc. An ionised outflow is also present in this system \citep{Crenshaw2000}.
In the only non-local study of a QSO/SB system with both molecular (sub-kpc scale) and ionised (kpc-scale) AGN- driven winds (i.e. SDSSJ1356, z=0.123; \citealt{Greene2012}), the gas outflows in the two components have different locations, velocities, and morphologies \citep{Sun2014}, which points towards a complex interplay of the two phenomena, or the presence of different outflow episodes, or a combination of the two. 

From the point of view of the overall gas content, observations of cold molecular gas reservoirs at high redshift (z$>1$; see \citealt{CW2013}) have been crucial in studying consumption rate and  excitation state in normal galaxies (e.g. \citealt{Tacconi2013,Genzel2015,Sargent2014,Tacconi2018}), submillimetre galaxies (SMGs; \citealt{Bothwell2013}), and quasar systems (e.g. \citealt{Solomon2005,Riechers2011,Bothwell2013,Feruglio2014}). In particular, spectral line energy distributions (SLEDs) and excitation modelling studies of $^{12}$CO line fluxes up to mid- or high-J transitions of high-z unobscured QSO hosts (e.g. \citealt{Gallerani2014}) have suggested that the molecular gas is in a higher excitation state than  in SMGs; this agrees with the picture that unobscured quasars represent a subsequent stage in the early evolution of massive galaxies (e.g. \citealt{Hopkins2008,Aravena2008,Riechers2011_QSOz3}). The fact that high star formation efficiencies (SFE; defined as the efficiency with which gas is converted into stars, e.g. star formation rate (SFR) over M$_{\rm gas}$) are also observed in these systems (e.g. \citealt{Solomon2005,Riechers2011}) seems to be instead in contradiction with model predictions, according to which the SFR is expected to have already  been substantially diminished. A possible explanation may be that, instead of higher efficiency in converting gas into stars, the high SFE observed in very luminous QSOs is a result of a low molecular gas content with respect to their current SFR, as indeed is expected if most of the gas has been already consumed.

Molecular gas studies at z$>$1 have been extended only recently to other classes of AGN systems (e.g. \citealt{Kakkad2017}), including reddened systems (e.g. \citealt{Banerji2017}), and/or to QSOs with pre-existing evidence of outflow signatures in the ionised gas component extending over the entire host galaxy \citep{Polletta2011,B15_PDBI,Carniani2017,Popping2017}. In several cases, the millimetre observations returned only upper limits for the CO flux, suggesting that the available samples used to predict the CO luminosities (unobscured QSOs, mostly associated with ULIRGs systems) were not representative of the molecular gas content of the full AGN population.

In this paper we present sensitive ALMA observations of the dust continuum and CO(5-4) transition in an obscured QSO at z$\sim$1.6. The target is XID2028 (z=1.5930), originally discovered in the XMM-COSMOS survey \citep{Brusa2010}, which is the  archetypical object in the feedback phase. The presence of a massive outflow in the ionised gas component of XID2028, traced by the [O III]5007 emission, has been independently confirmed by X-shooter slit spectroscopy \citep{Brusa2015,Perna2015} and SINFONI IFU observations \citep{Cresci2015}. In fact, XID2028 hosts one of the most massive (M$_{\rm ion}> $250 M$_\odot$ yr$^{-1}$ with v$>$1500 km s$^{-1}$) and most extended (out to scales of $\sim$13 kpc) outflows detected in a high-z QSO. Most importantly, the outflow lies exactly in the centre of a cavity in star forming regions in the host galaxy, as traced by the narrow H$\alpha$ emission line map at $\sim$0.5\arcsec\ resolution and rest-frame U-band imaging; this suggests that the wind is removing the gas from the host galaxy (negative feedback), and at the same time is triggering star formation (SF) by outflow-induced pressure at the edges (positive feedback; see also \citealt{Silk2013,Cresci2015_MUSE}). For these reasons, XID2028 can be considered one of the best targets to perform studies aimed at searching for and mapping an outflow component in the molecular gas phase at z$>$1.  

We observed XID2028 with the PdBI interferometer in the CO(3-2) transition. In \citet{B15_PDBI} we detected line emission at 5.3$\sigma$ with a beam of 4.5-3.5\arcsec\ and we inferred a value for the molecular gas mass in the range M$_{\rm gas}$ = 2.1-9.5$\times10^{10}$ M$_\odot$, depending on the assumed CO-H$_2$ conversion factor\footnote{The CO-H$_2$ conversion factor is defined as $\alpha_{\rm CO}$ = M(H$_2$)/L'$_{\rm CO}$, where M(H$_2$) is the H$_2$ mass in M$_\odot$ and L'$_{\rm CO}$ the CO(1-0) line luminosity in K km s$^{-1}$ pc$^2$ \citep{Bolatto2013}. The units  M$_\odot$/(K km s$^{-1}$ pc$^2$ are omitted in the rest of the text.} $\alpha_{\rm CO}$ (from $\alpha_{\rm CO}$=0.8 to $\alpha_{\rm CO}$=3.6). When coupled with the measured stellar mass, this translates into a molecular gas fraction $\mu_{\rm mol}$ =5-20\%. This low molecular gas fraction, when compared to that expected for M$_\star> 10^{11}$ M$_\odot$ systems with the same observed specific SFR, is consistent with a scenario in which the cold gas in the host galaxy of XID2028 has been already partly depleted by the effects of the strong QSO feedback \citep{B15_PDBI}. The regions where SF is still ongoing may be the clumpy gas reservoirs located at the edge of the outflow cavity and seen in the narrow H$\alpha$ map \citep{Cresci2015}. Given that the narrow H$\alpha$ map is only sensitive to regions of unobscured SF, a full census of the molecular gas and associated SF regions is needed to test our positive feedback scenario. Moreover, the SINFONI data probe only the atomic ionised gas component. Detection of a molecular outflow and a measurement of the associated mass outflow rate is needed to estimate the true energetics of the outflow process and its impact on the host galaxy. These two reasons motivated our ALMA follow-up observations for dust continuum and high-resolution imaging of CO transitions.

The paper is organised as follows: Section 2 presents the ALMA observations and data analysis. Section 3 presents the optical and near-infrared photometry (including data obtained in the commissioning of ARGOS at LBT) in comparison with the ALMA data. 
Sections 4, 5, and 6 discuss the results, and in particular the kinematics as inferred from the bulk of the molecular gas; the gas consumption of XID2028; and the detection of the molecular outflow. Finally, Section 7  summarises our conclusions. Throughout the paper, we adopt the cosmological parameters H$_0$ = 70 km s$^{-1}$ Mpc$^{-1}$ , $\Omega_{\rm m}$ =0.3, and $\Omega_\Lambda$=0.7 \citep{Spergel2003}. In quoting magnitudes, the AB system is used unless otherwise stated. We adopt a Chabrier initial mass function to derive stellar masses and SFRs for the target and comparison samples. 
The physical scale is 1\arcsec$\sim$8.5 kpc at the redshift of the source.

   \begin{figure}[!t]
   \centering
 \includegraphics[width=8.0cm,angle=0]{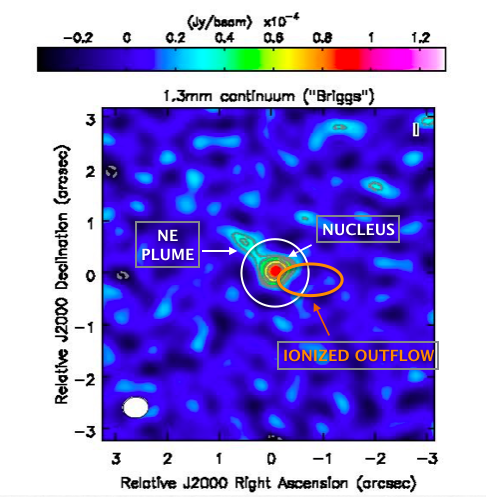}
\footnotesize
      \caption{1.3 mm (Band 6) continuum image of XID2028 obtained with Briggs weighting. 
      Dashed contours are drawn at negative $2\sigma $  fluctuations. Solid contours are drawn at 3, 4, 6, and $8\sigma $ with respect to the image rms (with this weighting scheme, rms=9.6$\mu $Jy beam$^{-1}$). The large FoV is shown to highlight the significance of the continuum detection. The two components detected in the continuum (`nucleus' and `plume') are labelled in the image. The (white) circle represents a reference region of 1\ensuremath  {^{\prime \prime }}\ diameter centred on the continuum peak emission. The orange ellipse denotes the position of the ionised outflow reported in Cresci et al. (2015a; see also Section 6). North is up and east is left.
              }
         \label{continuum}
   \end{figure}
%

\section{ALMA observations of XID2028}

XID2028 was observed during ALMA Cycle 3 with the 12 m array in Band 6 for a total time on source of 3.5 hrs.
The array included 38 antennas with a maximum baseline of 704 m. 
The phase centre of the dataset was set to the Hubble Space Telescope (HST) position of the QSO nucleus (RA=10:02:11.29, DEC=+01:37:06.67). The primary beam, corresponding to the field of view (FoV) of the observation, is $\sim$22\arcsec. The conditions were overall good and only a small fraction of the data ($\sim$15\%) was excluded due to standard pipeline processing flagging.

The spectral set-up covers a total bandwidth of 7.5 GHz. For two spectral windows (SPWs) the correlator was set to frequency division mode (FDM) with a bandwidth of 1875 MHz and a channel spacing of 3.9 MHz corresponding to a velocity resolution of $\sim$5 km/s, while two additional SPWs with 2 GHz bandwidth (time division mode; TDM) were used for continuum measurements. The two FDM SPWs were centred at 222.24 GHz, which is the frequency expected for the CO J=5-4 transition ($\nu_{\rm rest}$=576.268 GHz; hereafter: CO(5-4)) at the source redshift, and at $\sim$239.5 GHz to observe the HCN(7-6) and HCO$^+$ (7-6) transitions with $\nu_{\rm rest}$=620.304 GHz and $\nu_{\rm rest}$=624.208 GHz, respectively.

The data were calibrated using the ALMA pipeline. The quasar J1058+0133 (with a flux of 3 Jy at 222.24 GHz) was used for bandpass and absolute flux calibration, which yields an absolute flux accuracy of about 10\% at the observed frequency. The quasar J0948+0022 was used as gain calibrator. Images were created using the common astronomy software applications package (CASA v4.7; \citealt{McMullin2007}).

   \begin{figure*}[!t]
   \centering
 \includegraphics[width=8.7cm,angle=0]{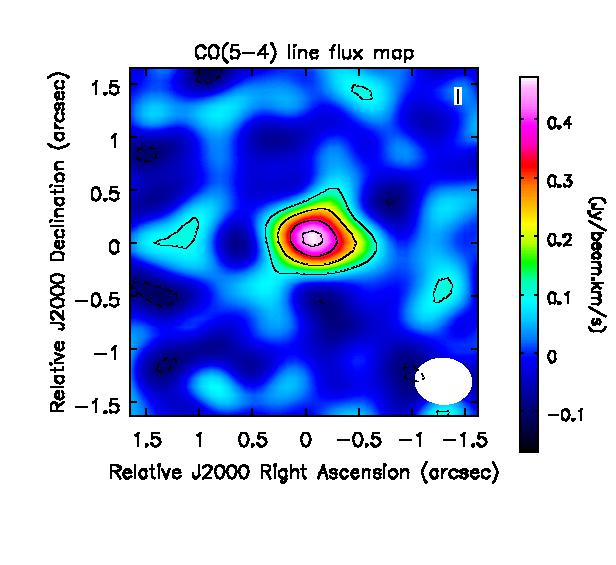}
 \includegraphics[width=8.7cm,angle=0]{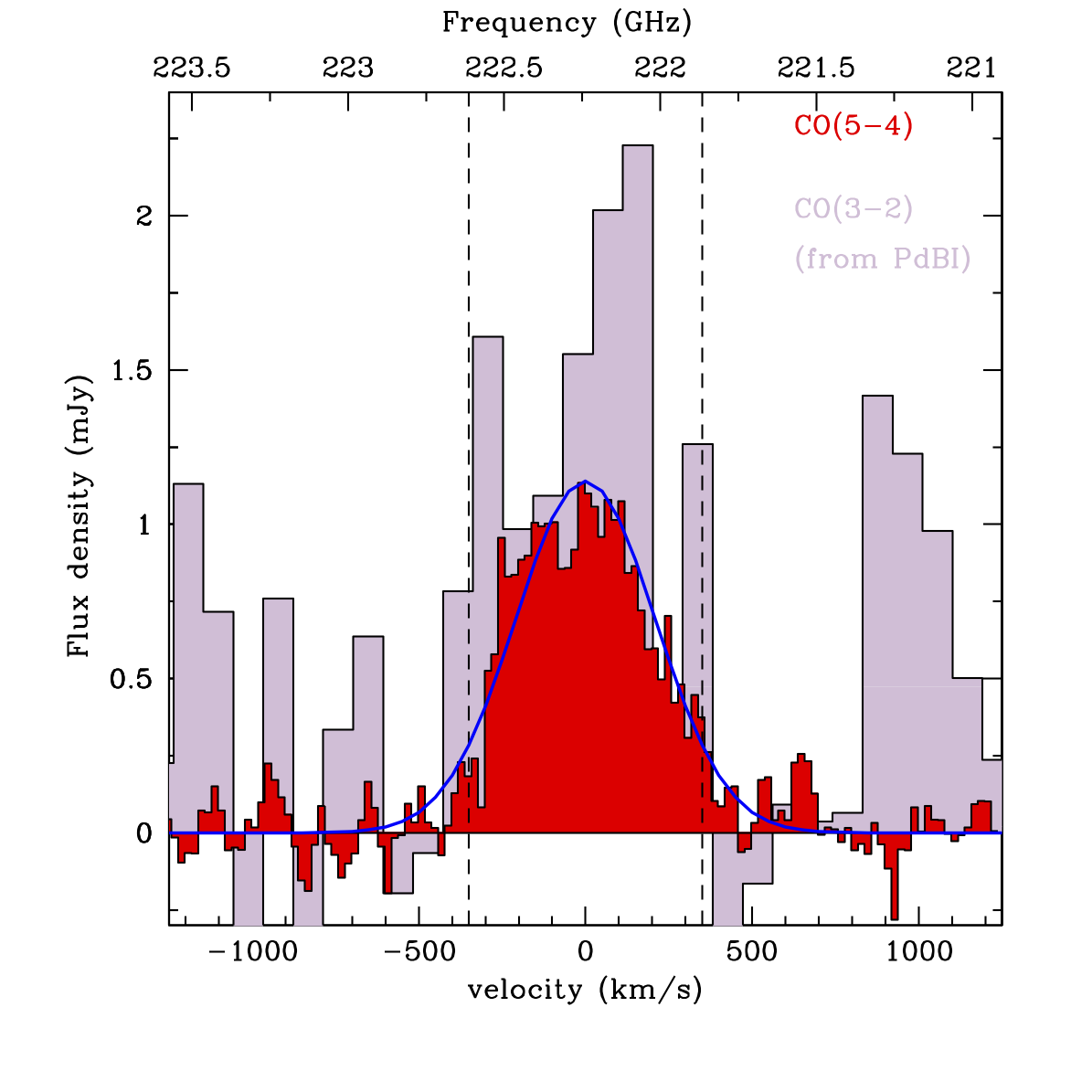}
\vspace{-1cm}
\footnotesize
      \caption{{\it  Left panel}: CO(5-4) line image integrating the emisson from -800 to 800 km s$^{-1}$ (see Section 2.2), imaged with the Briggs weighting scheme. Contours are drawn at -3 (dashed), 3, 6, 10, and 14$\sigma $ (solid; $\sigma $=0.032 Jy beam$^{-1}$ km s$^{-1}$). The ellipse in the lower right corner denotes the beam size (0.54"$\times $0.43"). The FoV is about 3$\ensuremath  {^{\prime \prime }}\times 3\ensuremath  {^{\prime \prime }}$ ($\sim 25\times 25$ kpc; 1" corresponds to 8.5 kpc at the redshift of the source). The detected CO(5-4) emission is confined to the central $\sim $4x4 kpc. {\it  Right panel}: Continuum-subtracted ALMA spectrum extracted around the CO(5-4) transition, rebinned at 20 km s$^{-1}$ per channel. The red filled histogram shows the spectrum extracted from a polygonal aperture encompassing the 3$\sigma$ contours of the line emission shown in the left panel (roughly corresponding to the 1\arcsec X-shooter slit width; Perna et al. 2015). 
      The dashed lines at -350 and 350 km s$^{-1}$  are used to define the blue and red tails of the CO emission (see Section 6). The purple histogram reproduces the CO(3-2) emission from PdBI, taken from Brusa et al. (2015b).   The blue line shows the fit with a single Gaussian with FWHM=550 km s$^{-1}$.           }
         \label{Cospectrum}
   \end{figure*}
%

\subsection{1.3 mm continuum}

To estimate the dust continuum emission at 1.3 mm ($\sim500 \mu$m rest frame), we collapsed the line free channels in the two TDM spectral windows (in the ranges 224-225.5 GHz and 236.5-238.2 GHz). 

We first reconstructed a continuum image with the CASA task clean, weighting the visibilities with the Briggs scheme to maximise the spatial resolution 
\citep{Briggs1995}. The clean beam of the observation is $0.54\arcsec\times0.45\arcsec$ with a position angle (i.e. measured clockwise from the positive y-axis) of 86 degrees.
The continuum image of XID2028 is shown in Fig.~\ref{continuum} with the beam plotted in the lower left corner. 
The image illustrates that the QSO host galaxy is clearly resolved into a central source (`nucleus') and a fainter feature extending towards the north-east direction (`plume'). The plume does not have any obvious counterpart in the SINFONI line emission maps nor in the HST rest frame U band, but it may be connected with a faint source detected in our high-resolution near-infrared data (see Section 3).

We then estimated the continuum flux by fitting with a Gaussian function the visibilities of the continuum dataset (via the CASA task \texttt{uvmodelfit}). We measured a continuum flux of 142 $\mu$Jy for a detection significance of 19$\sigma$ when the visibility noise of 7.5 $\mu$Jy is considered. 
After deconvolving for the beam,  the Gaussian fit  returned a FWHM size of 0.30\arcsec$\pm0.05\arcsec$ for the continuum emitting region in the nucleus,  which translates into an effective radius $\sim1.3$ kpc.

Coupling the information on  the spatial extent observed in the dust continuum image 
and the SFR as inferred from the SED fit presented in Perna et al. 2015 ($\sim$270 M$_\odot$ yr$^{-1}$; see also Appendix A.1), we estimated a SFR surface density $\Sigma_{\rm SFR}\sim25^{+13}_{-17}$  M$_\odot$ yr$^{-1}$ kpc$^{-2}$. This SFR surface density is significantly higher than that observed in normal star forming galaxies at z$\sim$1.5 ($\Sigma_{\rm SFR}\sim1-5$  M$_\odot$ yr$^{-1}$ kpc$^{-2}$; \citealt{Daddi2015}) and more similar to those observed in bright submillimetre galaxies (e.g. $\Sigma_{\rm SFR}\sim 20$  M$_\odot$ yr$^{-1}$ kpc$^{-2}$ in GN20 presented in \citealt{Magdis2012a}; see also \citealt{Gilli2014,Hodge2016}).

\subsection{CO(5-4) emission line}

The main goal of the ALMA observation is the characterisation of the dense molecular gas reservoir via the study of the CO(5-4)  transition. The redshift of the narrow component of the rest-frame optical emission lines (z=1.5930) was adopted to convert the frequency to velocity space\footnote{The systemic redshift is based on the narrow components detected in the H$\alpha$, [N II] and [O III] lines seen in the X-shooter spectrum and it has an associated  error of $\Delta$z$\sim 0.0002$; see \citealt{Perna2015}}.

The left panel of Fig. 2 shows the continuum-subtracted flux map integrated over the velocity range [-800$\div$800 km s$^{-1}$] around the expected line frequency, limited to the inner 3\arcsec$\times$3\arcsec ($\sim25\times25$ kpc) centred in XID2028. Also in this case we reconstructed the CO(5-4) line image by weighting the visibilities with the Briggs scheme. The clean beam of the observation is $0.54\arcsec\times0.43\arcsec$.
The CO(5-4) line is clearly detected, and confined to the inner 1\arcsec$\times$1\arcsec area, corresponding to the circle in Fig.~\ref{continuum}. 

The right panel of Fig. 2 shows the line spectrum extracted from the continuum-subtracted velocity cube with 20 km s$^{-1}$ binning (red histogram), over a polygonal region encompassing the CO-emitting region (taken at the 3$\sigma$ contours in the left panel) and roughly corresponding to the 1\arcsec\ diameter shown in Fig.~\ref{continuum}.  The line is significant at $\sim30\sigma$.  The CO(3-2) spectrum obtained from PdBI \citep{B15_PDBI} is also plotted as a purple histogram. The comparison of the CO(3-2) and CO(5-4) emissions shows no velocity offsets and an overall similar line width: the fit with a single Gaussian component returns FWHM$\sim$500 km s$^{-1}$ (solid blue curve in the right panel of Fig.~2). However, the reduced  $\chi^2$ square is high (2.27), which is likely due to the presence of deviations between the data and the Gaussian modelling (e.g. at  $-$500 km s$^{-1} <v<-$150 km s$^{-1}$; see also Section 6 for a more detailed discussion).

We then estimated the line flux by fitting with a Gaussian function the visibilities of the continuum-subtracted dataset (using the CASA task \texttt{uvmodelfit}), in the same velocity range as above. We retrieve values for the total line flux (I$_{\rm CO(5-4)}$ = 0.77$\pm$0.032 Jy km s$^{-1}$), the spatial extent, (FWHM=0.33$\pm$0.02\arcsec) and the line centroid (RA=10:02:11.28, DEC=01:37:06.64). The CO(5-4)  and the dust continuum peak emission  originate from the same region, at the phase centre position.

Following \citet{Solomon1997}, the measured CO(5-4) flux translates into a line luminosity of 
logL'$_{\rm CO(5-4)}$/(K km s$^{-1}$ pc$^2$)=9.63$\pm0.05$. 
From the observed far-infrared luminosity associated with the starburst (SB) component (logL$_{\rm IR}$ =12.47 erg s$^{-1}$; integrated from 8 to 1000 $\mu$m), well constrained from the PACS/SPIRE dataset, following the analysis of L'$_{\rm CO(5-4)}$-L$_{\rm IR}$ correlation in \citet{Daddi2015} we would expect logL'$_{\rm CO(5-4),L_{FIR}}$ =9.95 K km s$^{-1}$ pc$^2$. Although the observed value of the CO(5-4) line luminosity is still consistent within 2$\sigma$ given the scatter of the correlation, it is in fact a factor of $\sim$2 lower than that expected on the basis of the SFR.

Finally, we analysed the FDM spectral window centred at $\sim$240 GHz an tuned to observe the HCN(7-6) and HCO$^+$(7-6) transitions. Both lines remain below the detection threshold and we can provide $\sim3\sigma$ upper limits of 0.07 Jy km s$^{-1}$ for both transitions (assuming a FWHM=500 km s$^{-1}$). 

Table 1 reports the coordinates XID2028, its redshift, and all the measurements performed in our ALMA dataset.

\begin{table}
        \centering
         \caption{ALMA measurements}
          \begin{tabular}{m{6cm}  r} 
                \hline\hline 
                R.A. (hms, J2000) & 10:02:11.28  \\
                Dec. (dms, J2000) & 01:37:06.64 \\
                $z_{\rm spec}$ & 1.5930\\
                S$_{\rm cont, nucleus}$ ($\mu$Jy) & 142$\pm7.5$\\ 
                 FWHM (deconvolved, \arcsec) & 0.30\arcsec$\pm$0.05\arcsec \\
                 r$_{e, dust}$ (kpc)  & 1.3$\pm0.2$ \\ 
                 $\nu_{\rm CO(5-4),obs}$ & 222.2398 \\                                          
                $I_{\rm CO(5-4)}$\,(Jy\,km\,s$^{-1}$)  & 0.77$\pm$0.03  \\  
                                 FWHM (deconvolved, \arcsec) & 0.33\arcsec$\pm$0.02\arcsec \\
                 r$_{e, CO}$ (kpc)  & 1.4$\pm0.1$ \\ 
                $\log L^{\prime}_{\rm CO(5-4)}/K\,km\,s^{-1}\,pc^{-2}$)  & 9.63$\pm0.05$ \\
                                \hline 
        \end{tabular}
\tablefoot{Rows description: 
1-2) Coordinates of the continuum and CO(5-4) line emission (J2000); 
3) spectroscopic redshift from X-shooter; 
4) continuum flux at 500 $\mu$m rest frame of the nucleus source; 
5) deconvolved FWHM size; 
6) effective radius of the dust continuum; 
7) observed frequency of the CO(5-4) transition; 
8) CO(5-4) velocity integrated line intensity over 1600 km s$^{-1}$; 
9) deconvolved FWHM size; 
10) effective radius of the CO(5-4) emitting region;
11) CO(5-4) line luminosity. }
\end{table}

\subsection{Other CO transitions from PdBI and ALMA}

In \citet{B15_PDBI} we reported for XID2028 a 3$\sigma$ upper limit for the CO(2-1) line from the PdBI observation at 3 mm of I$_{\rm CO(2-1)} <$0.53 Jy km s$^{-1}$. For the CO(3-2) transition, instead, we measured a flux in the 2 mm band of I$_{\rm CO(3-2)}$=1.23$\pm$0.25 Jy km s$^{-1}$. In both cases, the beam of the observation was around 4$\arcsec$. The centroid of the CO(3-2) detection showed an offset of $\sim1.5$\arcsec\ from the quasar position (see Fig. 2 in \citealt{B15_PDBI}).
Given their larger beams, the published CO(3-2) and CO(2-1) measurements could include a contribution from larger scales than the relatively compact CO(5-4) emission. However, this is unlikely to be the case as the size of the dust continuum emission in the Rayleigh-Jeans tail, which is directly
equivalent to the low-J CO emission (\citealt{Magdis2012a,Genzel2015}), is basically the same as that of the CO(5-4) emission; this is strong evidence that most of the low-excitation gas is also confined to this region. 

With an a priori knowledge of the position and (Gaussian deconvolved) FWHM size of the CO(5-4) emitting region, it is now possible to model the flux emission in the PdBI data using the intrinsic spatial parameters obtained from our fit to the high signal-to-noise (S/N) ALMA data, over the same velocity ranges and without allowing for free variations of the parameters. We performed the measurement in uv space, where the model of the source is convolved with the beam, thus returning reliable measurements despite the large difference in beam size between the ALMA and PdBI data. 
We obtained in this case only a marginally significant detection (2$\sigma$) of I$_{\rm CO(3-2)}$=0.7$\pm$0.35 Jy km s$^{-1}$, although it is consistent at about 1.2$\sigma$ with our previous measurement. 
The fact that a flux extraction in the PdBI CO(3-2) map with the parameters of the CO(5-4) detection returns
a $\sim$2-fold lower CO(3-2) flux is consistent with the older measurement being boosted by noise\footnote{See the presentation by S. Guilloteau available at the IRAM web page, pages 22-28: http://www.iram.fr/IRAMFR/IS/IS2010/presentations/guilloteau-noise-101006.pdf.} . 

XID2028 was observed in ALMA Band 3 as part of the programme 2015.1.00171.S (PI: E. Daddi) with a snapshot (4 min) observation, targeting the CO(2-1) transition.
Using the spatial parameters obtained from our fit to the CO(5-4) data, and again consistently accounting for the 1.1$\arcsec$ beam of this ALMA dataset, we report a detection at $\sim3\sigma$ with an integrated flux of I$_{\rm CO(2-1)}$=0.35$\pm$0.12 Jy km s$^{-1}$, which is consistent with the PdBI upper limit. XID2028 remains instead undetected in the continuum and has a 3 mm upper limit of 0.1 mJy (3$\sigma$).

Finally, XID2028 was also observed at 850 $\mu$m in ALMA Band 7, as part of the programme 2015.1.00137.S (PI: N. Scoville) with a snapshot (2 min) observation aimed at detecting continuum emission of COSMOS high-z galaxies. The source was detected at low significance (2.6$\sigma$) with a continuum flux S$_{\rm cont,850\mu m}$=0.85$\pm$0.32 mJy.

\subsection{Estimating the CO(1-0) flux} 

\begin{figure}
   \centering
 \includegraphics[width=8.9cm,angle=0]{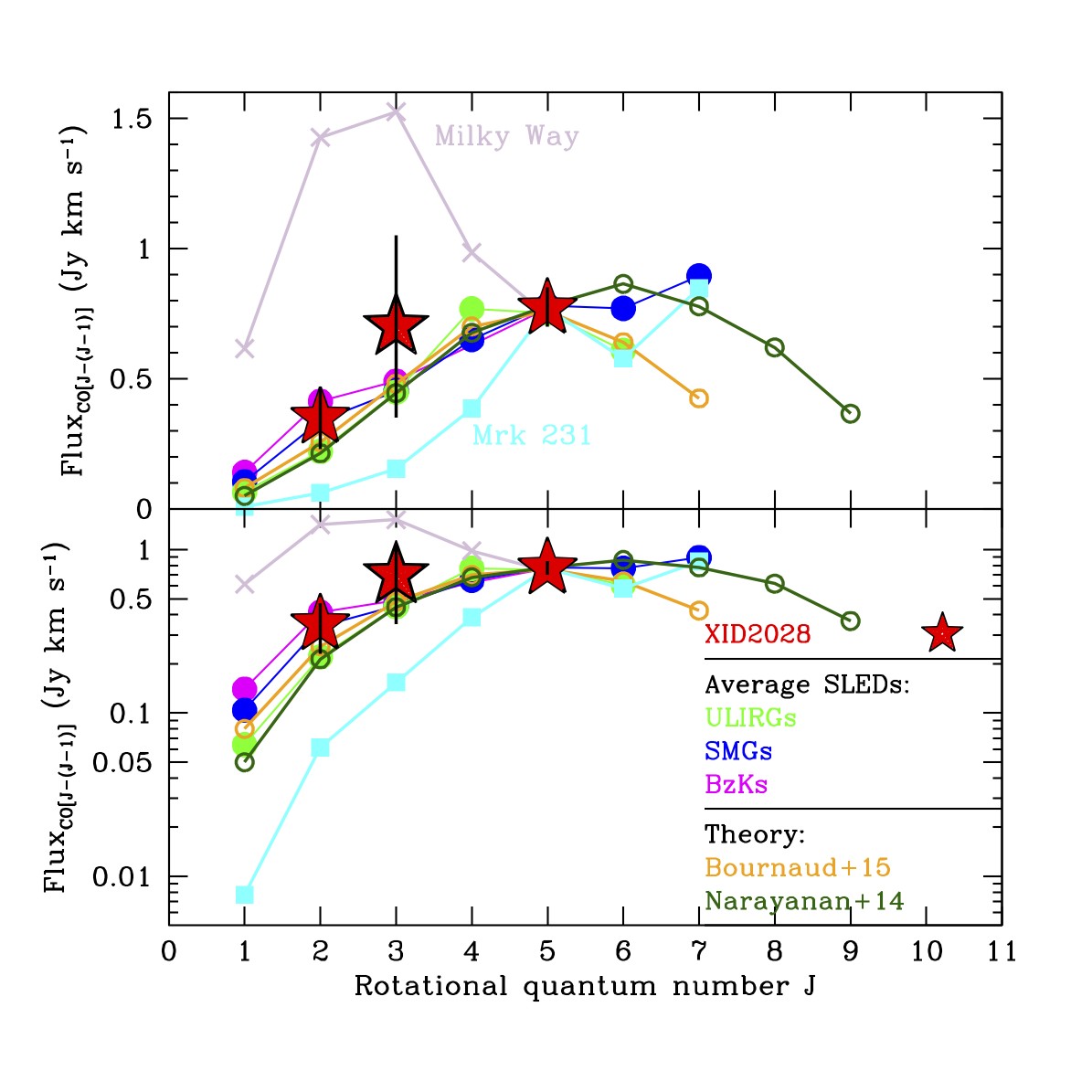}
      \caption{{\it Upper panel}: CO excitation ladder of XID2028 (stars) compared with average values obtained for various classes of sources (ULIRGs, SMGs, BzKs, and the Milky Way, as labelled; taken from Papadopoulos et al. 2012, Bothwell et al. 2013, Daddi et al. 2015, Fixsen et al. 1999), for single objects (Milky Way and Mrk 231; from  Fixsen et al. 1999 and van der Werf et al. 2010) and from simulations (Bournaud et al. 2015 and Narayanan\&Krumholz 2014). All SLEDs are normalised to the CO(5-4) flux observed in XID2028, for which we report an  uncertainty of 0.08 Jy km s$^{-1}$ also accounting for the 10\% flux calibration error.  {\it Lower panel}: Same as the upper panel, but with the y-axis in logarithmic scale, so that differences in the CO(1-0) flux extrapolations can be appreciated.}
         \label{COSLED}
   \end{figure}

In Fig.~\ref{COSLED} we plot the CO SLED for XID2028, i.e. the observed line flux in a given CO transition as a function of the upper $J$-level of the considered transition. The red stars are the following three measurements we have collected so far for XID2028: 
CO(2-1), CO(3-2), and CO(5-4) from ALMA and PdBI, as discussed in the previous subsections.
We compare the CO SLED of XID2028 with 
the average CO excitation ladders of ULIRGs \citep{Papadopoulos2012}, SMGs \citep{Bothwell2013}, and BzK galaxies \citep{Daddi2015}, colour-coded as labelled (see also \citealt{Dessauges2017}) and normalised to the CO(5-4) flux of our target. We also show the SLED of the Milky Way (purple crosses; \citealt{Fixsen1999}) and that of Mrk 231 (cyan crosses; from \citealt{Vanderwerf2010}), a SB-QSO system of which XID2028 was thought to be the high-z analogue (see \citealt{Brusa2010}).

At first sight, the CO SLED of XID2028 is consistent with the average SLEDs of other classes of sources that could be considered similar to this bright MS AGN (e.g. BzK are mostly MS galaxies at z$\sim$1.5-2; local ULIRGs often host luminous AGN).  
However, we note that the CO(1-0) flux extrapolated from the CO(5-4) measurement varies within a factor of 3, depending on the assumed SLED (lower panel of Fig.~\ref{COSLED}) in the range\footnote{From here on, we do not quote the additional error on the CO(1-0) flux related to the uncertainty on the CO(5-4) flux measurement and the flux calibration (overall $\sim10$\%), given that it is negligible with respect to the error associated with the assumptions on the SLED used to extrapolate the CO(1-0) flux}  I$_{\rm CO(1-0)}=$0.05-0.15 Jy km s$^{-1}$.

The CO SLED of XID2028 is instead not consistent with the SLED of the Milky Way and Mrk 231. 
The deviation from the Milky Way is expected given the different properties of the system. 
With respect to Mrk 231, in XID2028, we detect considerably higher CO(2-1) and CO(3-2) fluxes when compared to that observed in CO(5- 4). This may suggest that in XID2028 there is an extra cold gas component not present in Mrk 231, which may contribute to the flux at lower-J.

On the basis of numerical simulations of disc galaxies and mergers with molecular line radiative transfer calculations, \citet{Narayanan2014} have shown that, while the CO excitation ladders may scatter by more than one order of magnitude even in objects with similar SFR and stellar mass properties, they are instead predictable on the basis of the observed mean SFR surface density ($\Sigma_{\rm SFR}$). We show as green open circles the intensity ratios predicted following Equation 19 of \citet{Narayanan2014}, adopting $\Sigma_{\rm SFR}$=25 M$_\odot$ yr$^{-1}$ kpc$^{-2}$ (see end of Section 2.1). We also show the CO SLED expected from a typical SB-merger system in hydrodynamical simulations (orange open circles) presented by \citet{Bournaud2015}.
At low-J transitions, the two model predictions are basically indistinguishable and in broad agreement with the average CO SLEDs drawn from observations of various classes of high-z objects and with the CO SLED of XID2028. 
We note that the model by \citet{Narayanan2014} predicts the lowest CO(1-0) flux among all the average SLEDs plotted in Fig. 8 (I$_{\rm CO(1-0)}$ $\sim$0.05 Jy km s$^{-1}$). At higher J (J$> $5), instead, the two simulations predict remarkable different CO line fluxes, which can be tested in XID2028 with follow-up observations.

Given that in the \citet{Narayanan2014} modelling the only parameter that shapes the SLED is the $\Sigma_{\rm SFR}$, and this is independent of an AGN presence and its properties, the agreement observed with the model predictions suggests that the gas excitation does not change in the presence of a bright AGN, at least at low values of J. Indeed, emission from X-ray dominated regions (XDR) contribute to the observed SLEDs usually at higher CO transitions (see e.g. J1148 presented in \citet{Gallerani2014}).  This can also be tested with future ALMA or NOEMA observations. 

Extrapolating the CO(1-0) flux from the CO(5-4) emission, we have estimated the molecular gas mass associated with the component that produces this emission and within the region sampled by the CO(5-4) data, i.e. the inner few kpc area. With the ALMA CO(5-4) data alone we cannot therefore exclude the presence of less excited diffuse gas (not seen in CO(5-4)) with lower SF efficiency (not seen in dust continuum) on scales between 0.4\arcsec-1\arcsec (the host galaxy size; see Section 3). However, we have a stringent upper limit from PdBI for the CO(2-1) transition of 0.53 Jy km s$^{-1}$ \citep{B15_PDBI}, which refers to the emission of low-excitation gas over a spatial scale of $\sim4$\arcsec, much larger than the host galaxy scale. If we extrapolate the CO(1-0) flux from the PdBI CO(2-1) upper limit, we obtain an upper limit of 
I$_{\rm CO(1-0), 4\arcsec}<0.18$ Jy km s$^{-1}$ (using the BzK SLED, i.e. the SLED that returns the highest CO(1-0) flux among all the SLEDs considered above), very close to our quoted range from the CO(5-4) extrapolation (0.05-0.15 Jy km s$^{-1}$). 
We can therefore safely conclude that the majority of the gas is located in the central region and any possible contribution at larger scales is $<20-30$\%.

   \begin{figure}
   \centering
 \includegraphics[width=8.7cm,angle=0]{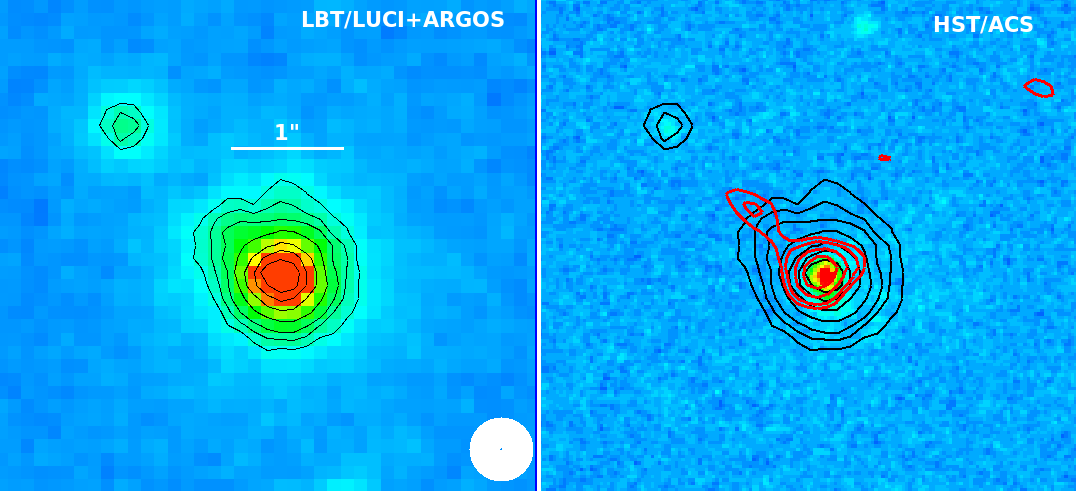}
      \caption{LUCI+ARGOS image (K band; left) and HST image (Advanced Camera for Survey F814W filter; right, taken from Brusa et al. 2015a,b) of a region of $\sim 6.5\times 6.5\ensuremath  {^{\prime \prime }}$ around XID2028 ($\sim 55\times $55 kpc). In both panels, we show the contours from the LUCI+ARGOS data (black, arbitrary levels chosen to trace the whole K-band emission). The FWHM of the LUCI+ARGOS data is $\sim 0.27\ensuremath  {^{\prime \prime }}$ and shown in the lower right corner of the left panel. The emission from the host galaxy is clearly resolved. In the right panel, we also plot the contours of the ALMA continuum at 1.3 mm (rest-frame 500 $\mu$m; red solid curves, from the right panel of Fig.~1). North is up and east is left.   }
         \label{hst}
   \end{figure}
%

\section{LBT data and NIR imaging}
XID2028 was observed at the Large Binocular Telescope (LBT) with the infrared cameras LUCI1 and LUCI2 in binocular mode (LUCI), during the commissioning of the Advanced Rayleigh Guided Ground Layer Adaptive Optics (ARGOS; \citealt{Rabien2010}), the new multi-laser guide stars ground layer adaptive optics system at the LBT. The ARGOS system provides ground layer correction of both mirrors of the LBT, using a system of three Rayleigh beacons on each side. This kind of AO system takes into account the effects of the lower altitude turbulence, yielding an effective seeing improvement over a wide angle, corresponding to the 4$^{\prime}\times4^{\prime}$ of the LUCI detectors.

XID2028 was imaged with the Ks filter (central $\lambda\sim2.2 \mu$m) in the commissioning nights of March 10 and March 14 2017, using both sides of the binocular telescope. The total integration time was 66 minutes for each instrument, LUCI1 and LUCI2, the infrared cameras at each eye of the telescope using dithered exposures of 10\arcsec\ each and a pixel scales of $\sim$0.12\arcsec. The final combined LUCI1 and LUCI2 images were registered to correct for a small residual rotation between the two and the slightly different pixel scale and distortions using point sources in the field. The average PSF derived from Gaussian fits of stars around the QSO is $\sim0.27\arcsec$. 

The final combined image of XID2028 is shown in the left panel of Fig.~\ref{hst}, zoomed in at the position of XID2028 ($\sim$55$\times$55 kpc in size). As a comparison, in the right panel we show the HST image obtained with the Advanced Camera for Surveys (ACS) with superimposed the contours from the K-band data (black). We note that the host galaxy is resolved in the K-band image with emission from the old stellar population tracing the galaxy stellar mass detected up to 10 kpc and not revealed in the HST data, which is mostly sensitive to the unobscured younger stars and the QSO continuum.

The red contours in the right panel of Fig.~\ref{hst} are taken from the 1.3 mm continuum map. The dust continuum peak is well aligned with the HST peak emission with only a marginal shift of $\sim$0.1\arcsec, which is well within the uncertainties of ALMA astrometry and is consistent with other shifts reported in the literature (e.g. \citealt{Decarli2017,Popping2017,Cibinel2017}). Therefore, the bulk of the CO emission, dust emission, and nuclear emission as traced by HST and LUCI all come from the same central region, within the limits of our observations ($\sim$0.1\arcsec, $\sim$850 pc). In the following analysis, we registered the ALMA, HST, and LUCI images so that all their emission peaks coincide.

Single component model fits to the LUCI image performed with GALFIT \citep{GALFIT} using Moffat, Sersic, or exponential profiles returned very poor results and bad residuals.
After the subtraction of the central point source with a PSF derived from the surrounding stars, we inferred a resolved host galaxy contribution to the total flux between 20\% and 50\%. From the GALFIT fit we also measured an effective radius r$_{e}\sim4.5$ kpc, assuming an exponential profile.  The residuals of the fit with these two components still reveal a clumpy structure slightly elongated in the NE direction (already recognisable in the unsubtracted image, see Fig.~\ref{hst}). 
We therefore tried a  three-components model (a Moffat function to fit the nucleus, a Sersic function to fit the halo, and an off-centre Moffat function that accounts for the NE asymmetry). 
From this last fit we inferred  a K-band magnitude of the nucleus Ks=18.51 (overall consistent with the value already available from the COSMOS photometry, Ks=18.68 from \citealt{Laigle2016}, when differences in the filters are taken into account),  a resolved host galaxy contribution to the total flux of the order of 20\% (with r$_e\sim10$ kpc), and a magnitude for the off-centre component of K$_s\sim20.5$ located at d = 0.63\arcsec ($\sim5.3$ kpc) in projection.

The off-centre component is detected at about the position of the dust continuum plume in the NE direction, and it has no counterpart in the rest-frame U band traced by HST. The continuum plume may  possibly trace the emission from a  heavily obscured companion to the host galaxy of XID2028 or from a tidal tail. Moreover, at $\sim2\arcsec$ ($\sim$17 kpc) in the same direction,  there is another point-like K-band detected object with no line emission detectable in our SINFONI or ALMA data cubes. \citet{Laigle2016} report for this source a photometric redshift estimate z$_{phot}$=1.6475, with a lower limit at the 68\% level of z$_{phot}$=1.5818, therefore consistent with the redshift of XID2028. However, the lack of spectroscopic confirmation for the K-band point-like source prevents a more quantitative analysis on the possible interaction between this object, nucleus of XID2028, and the dust continuum plume. 

\begin{table}
        \centering
         \caption{Derived quantities and physical properties}
          \begin{tabular}{m{6cm}  r} 
                \hline\hline 
                $\log(M_{\star}/M_\odot)$ & 11.65$^{+0.35}_{-0.35}$  \\
                $\log(L_{\rm IR}/L_\odot)$  & 12.47$^{+0.01}_{-0.05}$ \\
                SFR$_{\rm IR}$\,($M_\odot$ yr$^{-1}$)  & 270$^{+10}_{-30}$ \\
                v$_{\rm OF, ion}$ (km s$^{-1}$) & $\sim$1500 \\
                $\dot{M}_{\rm (OF, ion)}$ (M$_\odot$ yr$^{-1}$) & $>$300 \\
                $\dot{M}_{\rm (OF, neut)}$ (M$_\odot$ yr$^{-1}$) & $>$80 \\
   \hline
                $\Sigma_{\rm SFR}$ (M$_\odot$ yr$^{-1}$ kpc$^{-2}$) & $25^{+13}_{-17}$  \\
                $\log L^{\prime}_{\rm CO(1-0)}/K\,km\,s^{-1}\,pc^{-2}$)  & 9.84--10.31 \\
                T$_{\rm dust}$ (K) & 52$\pm5$ \\
                M$_{\rm ISM}$ (10$^{10}$ M$_\odot$) & 1.7--2.4 \\
                M$_{\rm mol, CO}$\,($10^{10} M_\odot$) & 1.1$\pm$0.5 \\                 
                $\mu_{\rm mol}$ & $<5$\%  \\            
                 t$_{\rm depl}$ (Myr) & 40--75 \\
                v$_{\rm OF, mol}$ (km s$^{-1}$) & $\sim$700 \\
                $\dot{M}_{\rm (OF, mol)}$ (M$_\odot$ yr$^{-1}$) & $\sim$50--350 \\
                                \hline 
        \end{tabular}
\tablefoot{Rows description: 
1) integrated stellar mass from SED fitting (see \citealt{Perna2015} and Appendix A.1); 
2) total IR luminosity derived through fitting of the IR component; 
3) integrated SFR derived from the total IR luminosity and applying the \citealt{Chabrier2003} calibration; 
4, 5) velocity of outflows and mass outflow rate inferred for the ionised outflow \citep{Cresci2015};
6) mass outflow rate of the neutral gas component \citep{Perna2015}; 
7) SFR surface density (see end of Sect. 2.1); 
8) extrapolated CO(1-0) luminosity from the observed SLED (see Sect. 2.4);
9) dust temperature inferred from the SED fitting (see Sect.~5); 
10) ISM mass derived from the dust continuum (see Sect.~5); 
11) molecular gas mass derived from the CO(1-0) line luminosity and assuming $\alpha_{\rm CO}$=0.8; (see Sect.~5); 
12--13) gas fraction and depletion timescale inferred for XID2028 (see Sect.~5); 
14--15)   outflow velocity and mass outflow rate in the molecular component (assuming $\alpha_{\rm CO}$=0.13--0.8). 
Measurements without errors and with quoted ranges are  dominated by systematic uncertainties (often of the order of 50\%) rather than statistical uncertainties. The quoted measurements/ranges should therefore be considered as order of magnitudes estimates of the relevant physical quantities. }
\end{table}

   \begin{figure}
   \centering
 \includegraphics[width=8.7cm,angle=0]{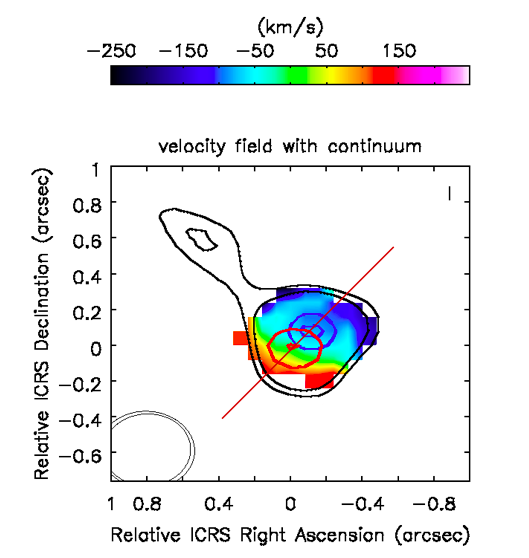}
      \caption{ CO(5-4) velocity map with the continuum superimposed (black contours, referring to the 3 and 4 $\sigma$ contour from Fig. 1) and with the centroids of the flux maps obtained by integrating the channels in the red (0$\div$300 km s$^{-1}$) and blue ($-$300$\div$0 km s$^{-1}$) part of the core line profile (red and blue contours, respectively; the contours are drawn at 5,6$\sigma$ and 8,9$\sigma$, respectively). The observed shift of the centroids is $\sim 0.1\ensuremath  {^{\prime \prime }}$. The red line shows the direction of the major axis of the rotating disc.}
         \label{vmap}
   \end{figure}
%

   \begin{figure}[!h]
   \centering
 \includegraphics[width=8.7cm,angle=0]{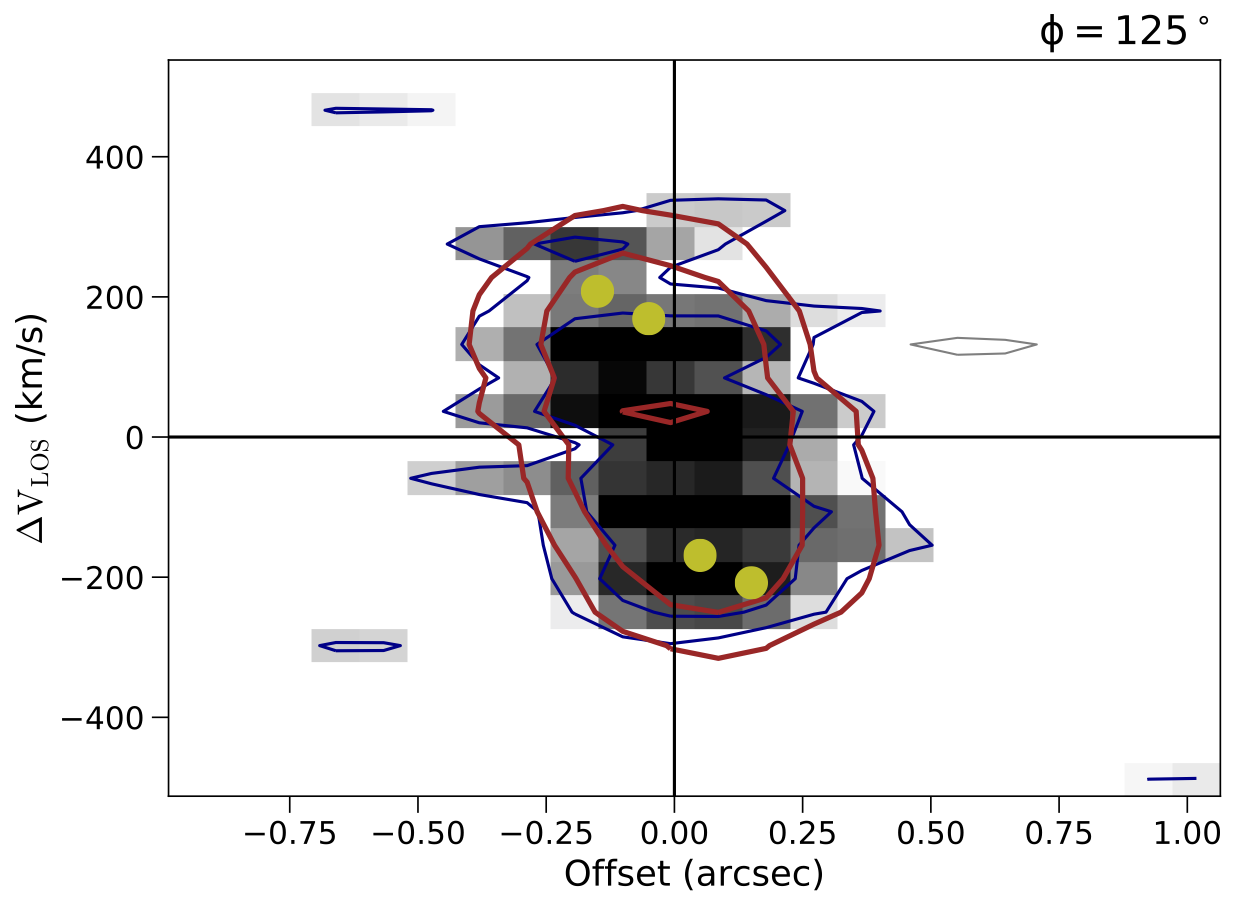}
      \caption{Position (i.e. offset along the major axis from the galaxy centre) vs. velocity diagram along the kinematic major axis of the molecular disc. The plotted velocity is along the line of sight (LOS). The blue and red contours show, respectively, the iso-density contours of the galaxy and the best-fit model found with $^{3D}$BAROLO, both starting from 2$\sigma$. The horizontal black solid line indicates the systemic velocity. The yellow circles denote the best-fit projected rotation velocity.}
         \label{PV}
   \end{figure}
%

\section{A fast rotating molecular disc in XID2028?}
Figure~\ref{vmap} shows the velocity field obtained with the CASA task \texttt{immoments}, showing the presence of an observed velocity gradient (from $\sim -$200 to $\sim$200 km s$^{-1}$).
The presence of a velocity gradient is also confirmed by integrating the CO data in the blue ($-$300$\div$0 km s$^{-1}$) and red (0$\div$+300 km s$^{-1}$) channels of the line core. The centroids of these detections are shown in Fig.~\ref{vmap} as blue and red contours to mark the blue and red line part, respectively. We measure a significant difference in the line centroids of these two flux maps of $\sim$1.5 pixels, which corresponds to $\sim$0.12$\pm$ 0.02\arcsec. A similar spatial scale clearly also emerges from the analysis of the position-velocity (PV) diagram (see below).

We fit the observed velocity and velocity dispersion field with the dynamical fitting routine of \citet{Cresci2009}.  In the hypothesis of a rotating disc as the origin for the observed velocity gradient,  we obtained a dynamical mass of $\sim6\times10^{10}$ M$_\odot$ at the scale sampled by the CO data, and we also retrieved an estimate of the inclination angle of $\sim$30 deg. A low inclination angle is consistent with the morphology of the host galaxy as suggested by the K-band data (see Fig.~\ref{hst}).  The zero velocity seems to agree very well with the continuum peak (shown as black contours in Fig.~5). This implies that the QSO nucleus is consistent with being located at the dynamical centre. We also infer a position angle for the molecular disc of about $\sim$125 deg with respect to the north-south direction.

We then used $^{3D}$BAROLO (3D-Based Analysis of Rotating Object via Line Observations), a tool for fitting 3D tilted-ring models to emission-line data cubes \citep{Barolo}. Assuming an inclination angle of 30 deg, we retrieved an intrinsic rotation velocity of $\sim$420 km s$^{-1}$. Although in this case we are also limited by the beam, the fit reproduces very well the emission in the PV diagram along the major axis of the molecular disc. Figure~\ref{PV} shows the PV diagram taken along
the major axis in grey scale and associated blue contours; the best-fit model from the 30 degrees inclination case is superimposed in red. The yellow circles denote the best-fit values for the observed velocity in the two rings considered in the fit by $^{3D}$BAROLO. 
The dynamical mass within the observed CO emission is M$_{\rm dyn}\sim8\times10^{10}$ M$_\odot$, which is comparable to our independent estimate.

The angular resolution of our data is not enough to unambiguously interpret the observed velocity gradient as due to disc rotation. As an alternative possibility, a velocity gradient and broad line profile could also be consistent with the presence of an ongoing major merger (see e.g. \citealt{Sharon2015, Decarli2017}). 
As mentioned in Sections 2.1 and 3, we detected a possible faint companion to XID2028 at $\sim0.6\arcsec$ to the NE direction from the nucleus (the plume and off-centre K-band source). However, 
if the plume were due to  tidal tail or a companion galaxy undergoing a merger, the velocity gradient should likely occur
along the direction connecting the QSO with the plume (south-west to north-east direction). 
Therefore, a rotating gas disc remains the most probable interpretation for the detected velocity gradient.

\section{Gas consumption in XID2028 and other obscured quasars as probe of feedback effects}

Following the prescriptions of \citet{Scoville2016} an estimate of the ISM mass can be obtained from the dust continuum flux, assuming a value for the dust temperature and a redshift-dependent coefficient for the dust opacity to take k-correction effects into account. 
We estimated the dust temperature (T$_{\rm dust}$) by applying the  multicomponent SED fitting code of \citet{Duras2017} to the multiwavelength (UV to Herschel/SPIRE bands) photometry from the COSMOS2015 catalogue of \citet{Laigle2016}, to which we also added the ALMA and PDBI data points in the (sub)millimetre regime (see Appendix A.1). 
We inferred a dust temperature T$_{\rm dust}$=52$\pm5$ K, which is significantly higher than that generally observed and assumed in normal main sequence (MS) galaxies at z$\sim$1.5 (T$_{\rm dust}\sim$25-30 K; \citealt{Magdis2012a,Santini2014,Magnelli2014}), and consistent with the values reported for high-z quasars and SMG (e.g. \citealt{Fu2012,Riechers2013,Gilli2014};  see discussion in \citealt{Duras2017}).
With this value for T$_{\rm dust}$ and the \citet{Scoville2016} scaling, the continuum flux\footnote{Applying the \citet{Scoville2016} recipe to the Band 7 data point, we derived a M$_{\rm ISM}$=3$\times$10$^{10}$ M$_\odot$. Given the lower detection significance of the ALMA Band 7 detection with respect to the ALMA Band 6 (2.6$\sigma$ versus $\sim$20$\sigma$), we adopted the value obtained from the Band 6 data for the ISM mass derived from the dust continuum.} at 1.3 mm measured at the position of the nucleus (S$_{\rm cont,1.3mm}$=0.142 mJy; see Section 2.1) translates into a gas mass M$_{\rm ISM}$=1.7$\times10^{10}$ M$_\odot$ (with a 10\% uncertainty from the Band 6 continuum flux measurement). We caution that most of the mass of the dust is likely to be  colder than the dust that emits most of the luminosity (see e.g. \citet{Scoville2017}, and  Mingozzi et al. in prep. for a local example). However, if we use a more typical value of T=25 K for the dust temperature, the gas mass increases only marginally (M$_{\rm ISM}$=2.4$\times10^{10}$ M$_\odot$).

   \begin{figure*}[!t]
   \centering
 \includegraphics[width=18cm,angle=0]{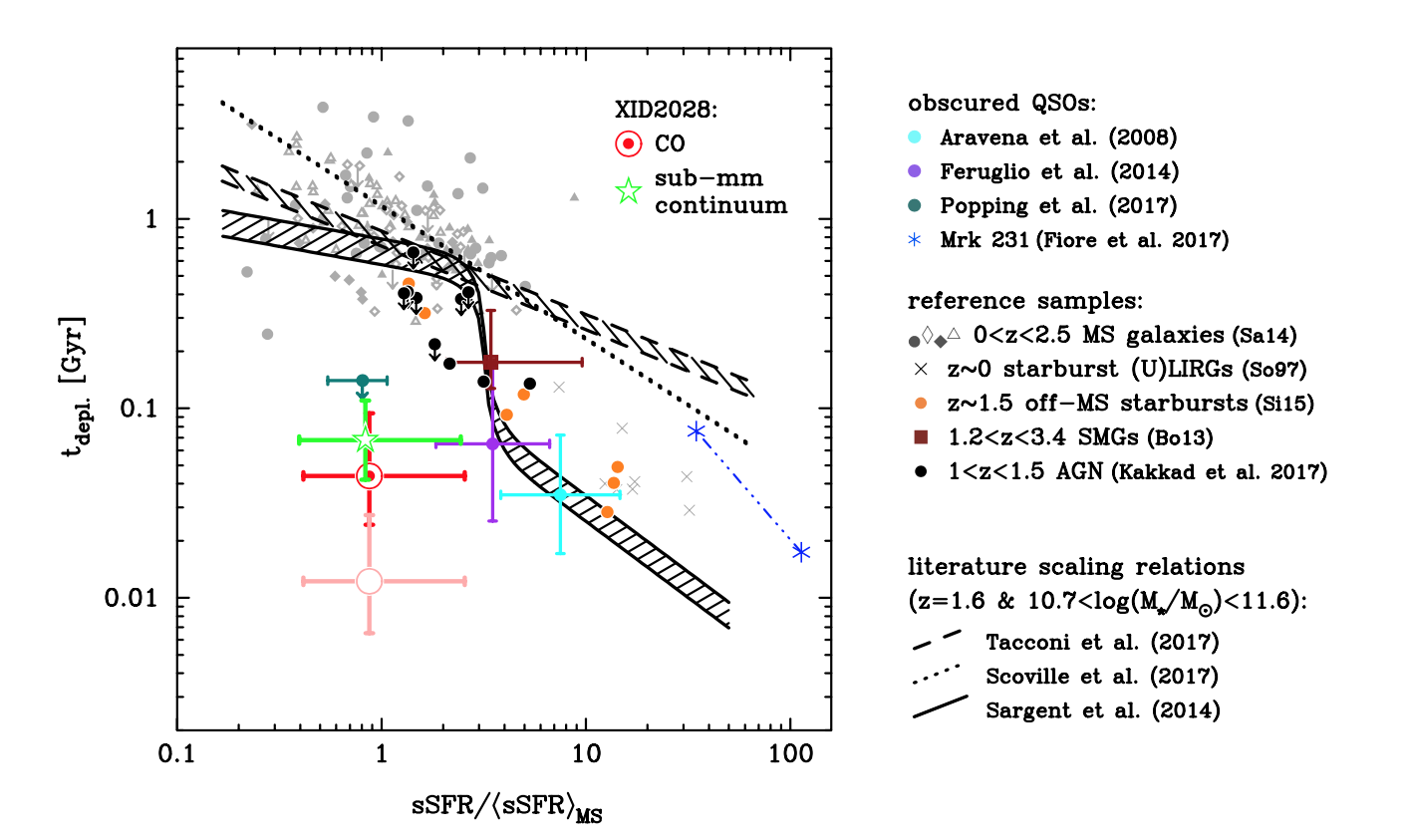}
\footnotesize
      \caption{Depletion timescale (M$_{\rm gas}$/SFR) plotted vs. the sSFR excess with respect to MS galaxies for obscured QSOs (coloured points) and reference samples (MS galaxies, SMGs/SBs, and AGN), as labelled. For XID2028 we report two values relative to the gas mass measurement obtained from the CO (red) and the submm continuum (green). For clarity, the values for XID2028 in the x-axis are slightly offset.  In addition, we also report the value of the depletion timescale obtained also taking gas consumption through the ejection by the AGN wind into account (pink point). The sSFR is normalised to the expected values for normal and SB galaxies predicted by the calibration presented in \citet{Sargent2014}. The black solid, dashed, and dotted line/regions traces the expected variation of the depletion timescale with the normalised sSFR for a MS galaxy at the redshift of XID2028, which has stellar mass in the range 5$\times 10^{10}-5\times 10^{11}$ M$_\odot $ predicted by the calibration presented in \citet{Sargent2014,Tacconi2018}, and \citet{Scoville2017}, respectively. XID2028 shows significantly lower depletion timescale than those expected for the properties of its host galaxy, lying a factor $\sim $10 to $\sim $20 below the black lines.  }
         \label{tdepl}
   \end{figure*}
%

In Section 2.4 we derived an estimate of the CO(1-0) flux in the range 0.05-0.15 Jy km s$^{-1}$.  This corresponds to a CO(1-0) luminosity in the range logL'$_{\rm CO}$(K km s$^{-1}$ pc$^{-2}$)=9.84-10.31. The compactness of the source detected in CO(5-4) and the relatively high dust temperature inferred from the SED grey body fit are typical of ULIRGs systems observed both locally and in the distant Universe. Therefore, we adopted a SB-like luminosity-to-gas-mass conversion factor, $\alpha_{\rm CO}$=0.8, in deriving the gas mass associated with the component that dominates the CO(5-4) emission. We inferred a gas mass of M$_{\rm gas,CO}\sim1.1\pm0.5\times10^{10}$ M$_\odot$, in fairly good agreement with our estimate from the Rayleigh-Jeans continuum ($\sim2\times10^{10}$ M$_\odot$).

We reported a stellar mass of 4.5$\times10^{11}$ M$_\odot$ for XID2028 (\citealt{Perna2015}; see also Appendix A.1). We calculated the molecular gas fraction, $\mu_{\rm mol}$, defined as the ratio of the molecular gas mass and the stellar mass ($\mu_{\rm mol}$=M$_{\rm mol}$/M$_{\star}$; see e.g. \citealt{Sargent2014,Genzel2015}), and we reported a value $\mu_{\rm mol}$$\lsimeq5$\%.
This value is notably smaller than those observed in high-z SMGs and quasars, which are associated with larger gas mass reservoirs when compared to the assembled stellar mass (see e.g. \citealt{Banerji2017}). These systems are indeed thought to be in an early stage of stellar mass assembly and are converting all the available molecular gas into stars with a very high efficiency. XID2028, instead, is hosted in a massive galaxy where most of the stars have already been formed and where dense molecular gas is present  only in a central, compact region. 

Finally, from the observed SFR and  gas mass, we can estimate the rate at which the gas is converted into stars, i.e. the gas depletion timescale (t$_{\rm depl}$=M$_{\rm gas}$/SFR). We inferred t$_{\rm depl} \sim40-75$ Myr using the CO and dust-fit derived gas masses, respectively. 
By comparing the gas depletion timescale and host galaxy properties of XID2028 to that of normal star-forming galaxies ($\sim0.5-1.5$ Gyr; \citealt{Sargent2014}), \citet{B15_PDBI}  proposed that, despite sitting on the MS, XID2028 is consuming its residual gas more rapidly than similar host galaxies at the same redshift. Our new analysis based on ALMA data further strengthens the significance of this result. 
The two values of the depletion timescale for XID2028 derived from the dust continuum and CO estimates in this work are plotted as red and green symbols in Fig.~\ref{tdepl} against the sSFR excess with respect to the MS.
The normalised sSFR values for XID2028 are based on our fiducial values of the host galaxies properties presented in Perna et al. (2015; see also Appendix A.1). 
The solid regions denote the median trend of t$_{\rm depl}$ with normalised sSFR, expected for galaxies in the stellar mass range $5\times10^{10}-5\times10^{11}$ M$_\odot$, at z=1.6, predicted by the calibration presented in \citet{Sargent2014}. XID2028 lies a factor $\sim10$ below this relation, while in \citet{B15_PDBI} the deviation was a factor of 2-3. The deviation is even more striking (a factor $\sim20$) when the calibrations presented in \citet{Tacconi2018} and \citet{Scoville2017} are considered (dashed and dotted regions, respectively).
The depletion timescale observed for XID2028 is instead more similar to that observed in bright SMGs  \citep{Bothwell2013} and off-MS SB galaxies  (off-MS; \citealt{Silverman2015}) at similar redshift.

In Fig.~\ref{tdepl} we also plot two bright obscured quasars (ULASJ1539 at z$\sim$2.5 from \citealt{Feruglio2014} and COSBO11 at z=1.8 from \citealt{Aravena2008}) characterised by short depletion timescales ($<100$ Myr), but overall  consistent with those expected given their high SFR ($>1000$ M$_\odot$ yr$^{-1}$). These systems have been proposed to be in the transition phase between a heavily obscured SB phase and the unobscured QSO phase. We complement these measurements for QSO and AGN systems, already presented in \citet{B15_PDBI}, with new data recently published regarding AGN at z$\sim1.5$. These data are 3D-HST GS30274 at z=2.23 (from \citealt{Popping2017}; a.k.a. K20 ID5 and GMASS 953, Talia et al. in preparation) and  the sample from \citet{Kakkad2017}, who presented CO(2-1) observations of 10 X-ray, selected AGN at z$\sim1-1.5$ in the COSMOS and CDFS fields, for which accurate estimates of SFR and M$_\star$ are available. Finally, we also plot the values for Mrk 231, considering the measurements reported at two different spatial scales as collected in \citet{Fiore2017}). 

The scarcity of observations of AGN at sSFR/sSFR$_{\rm MS}<$1 prevents a quantitative analysis. Indeed, the three detections reported in \citet{Kakkad2017} are in the range of sSFR/sSFR$_{\rm MS}$ = 2-6, which are more similar to those observed for the \citet{Aravena2008} and \citet{Feruglio2014} QSOs. For these bright FIR and submm sources, which have sSFRs comparable to SMGs, and for which we expect compact gas reservoirs, the short depletion timescales could be due to higher SFE in the galaxy (e.g. \citealt{Genzel2010,Daddi2010}). In fact, the gas fraction in the three CO detected \citet{Kakkad2017} AGN is $\sim$20-40\%, considerably higher than that observed for XID2028.

As already emerged from the SLED study, Mrk 231 is a luminous QSO with host galaxies properties and/or gas consumption mechanisms that are very different from XID2028. The only object similar to XID2028 is GMASS 953 presented in \citet{Popping2017}. Similar to XID2028, also evidence for ionised and molecular outflow has also been presented for GMASS 953 (Loiacono et al. in preparation). This suggests that, when powerful outflows are in place in AGN systems, the molecular gas reservoir may have been significantly affected, explaining the low depletion timescale observed. This scenario is also supported by short molecular depletion times found in local (e.g. \citealt{GarciaB2014,Casasola2015}) or higher-z \citep{Polletta2011} AGN with detected or potential outflows.  

\section{High velocity molecular gas tracing an extended molecular outflow}

   \begin{figure*}
   \centering
 \includegraphics[width=8.6cm,angle=0]{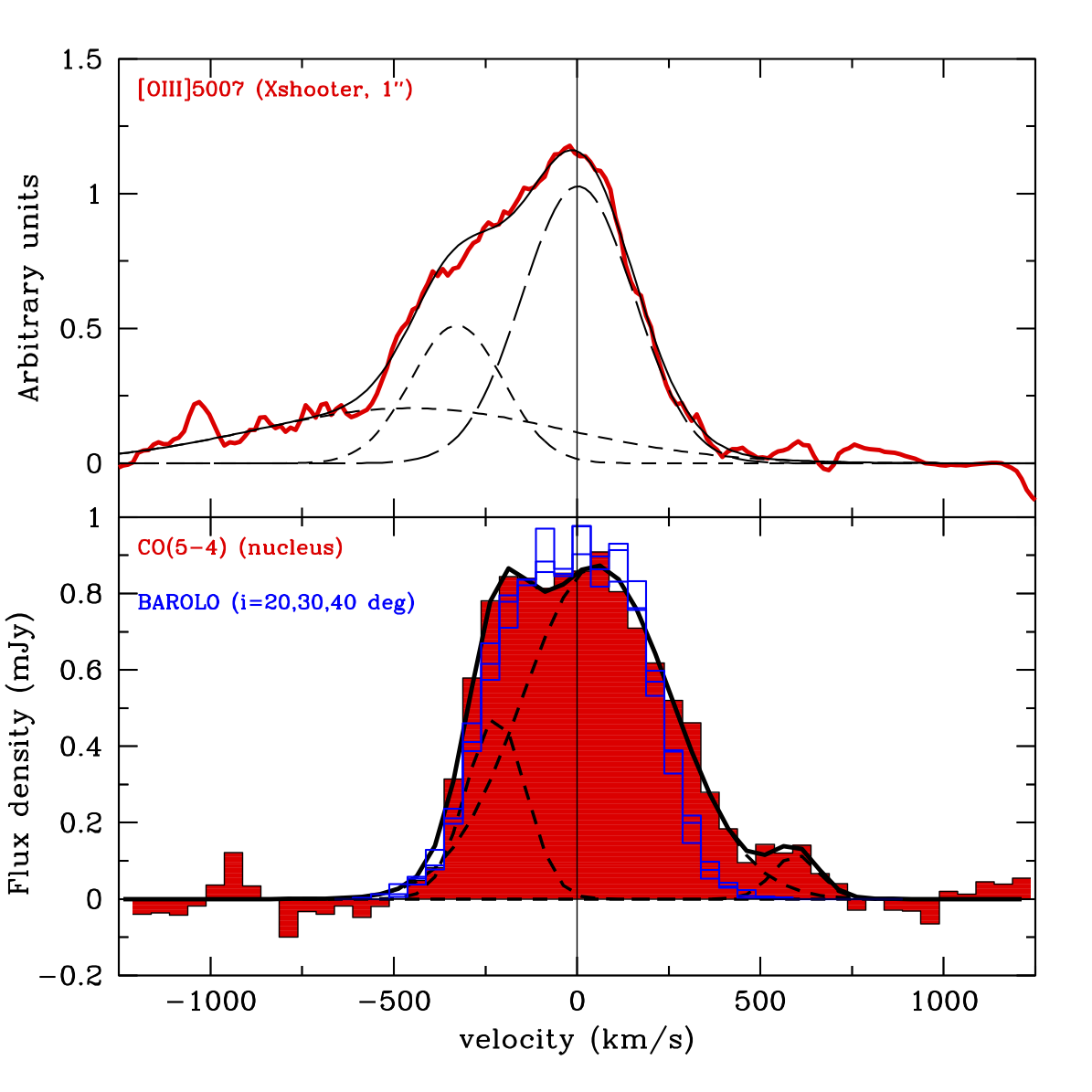}
 \includegraphics[width=8.8cm,angle=0]{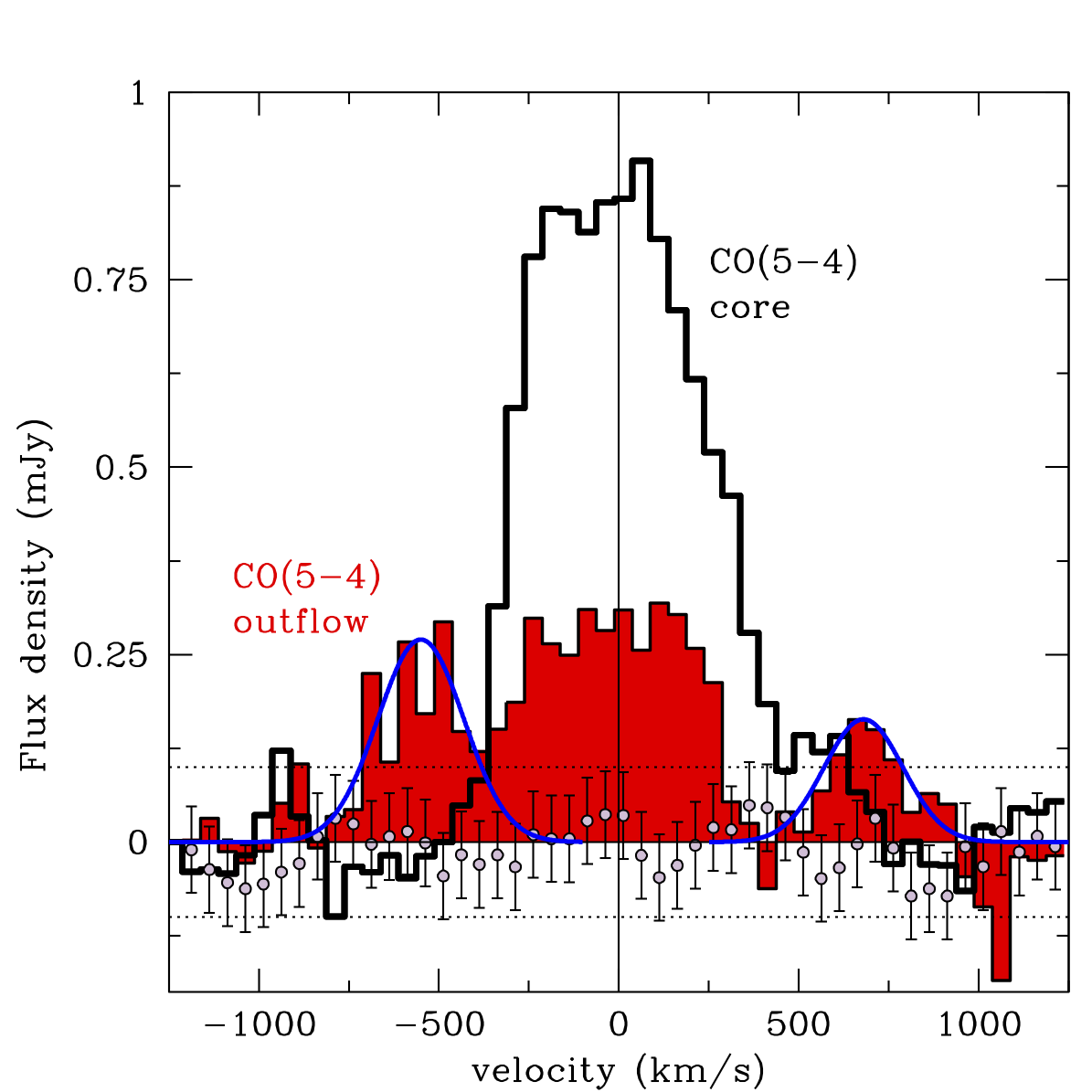}
\vspace{-0.2cm}
      \caption{{\it Left panel}: [O III] line profile (red curve taken from Perna et al. 2015; upper) and CO(5-4) line profile (red filled histogram; lower), both extracted from the $\sim$ 1\ensuremath  {^{\prime \prime }}\ diameter area shown in Fig.~\ref{continuum}. In contrast to Fig. 2, the continuum-subtracted CO(5-4) spectrum is binned at 50 km s$^{-1}$. The blue curves overplotted on the CO(5-4) data represent the spectral profile extracted from the model data cube obtained with $^{3D}$BAROLO (see Section 4),  assuming 20, 30, and 40 deg inclination.  The dashed lines represent the set of Gaussian components needed to reproduce the line profiles. {\it Right panel}: CO(5-4) spectrum extracted from a polygonal region encompassing the 1$\sigma$ contours shown in  the left panel of Fig~\ref{COtails} (red histogram extracted from the data cube obtained with a natural weighting scheme). We also show the $\pm$1$\sigma$ level  (dotted lines), the average of three noise spectra  taken randomly in the field over a region with the same area and shape as that of the outflow (purple circles), and, for reference, the CO(5-4) spectrum taken from the left panel as well (open histogram). The blue curves represent our Gaussian fit to the blue and red excesses at around $v\sim -600$ and $v\sim700$ km s$^{-1}$, respectively. 
      In all panels, the v=0 position is denoted by a solid vertical line.  }
         \label{Xshooter}
   \end{figure*}

%

   \begin{figure*}
   \centering
 \includegraphics[width=8.7cm,angle=0]{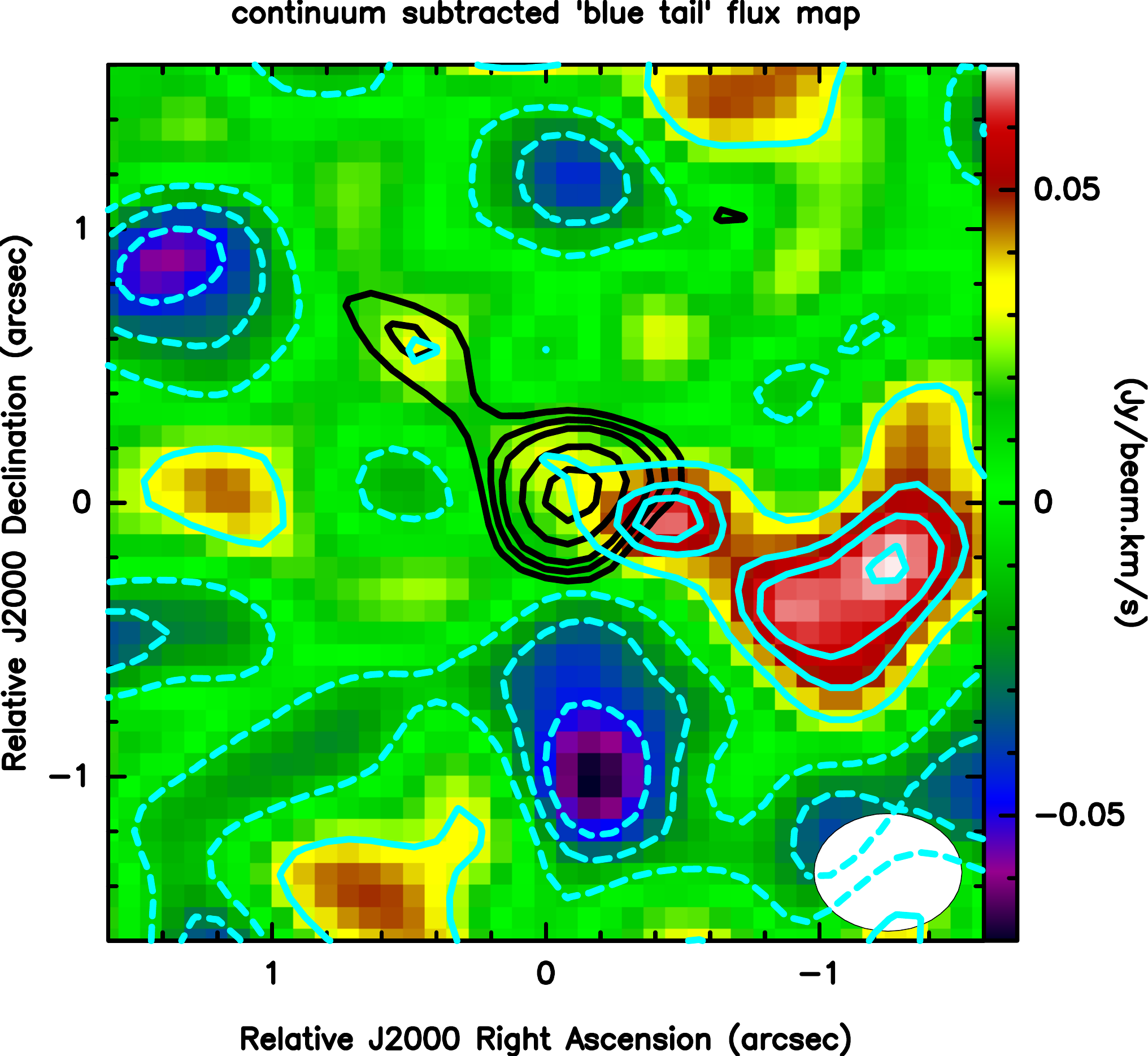}
 \includegraphics[width=8.7cm,angle=0]{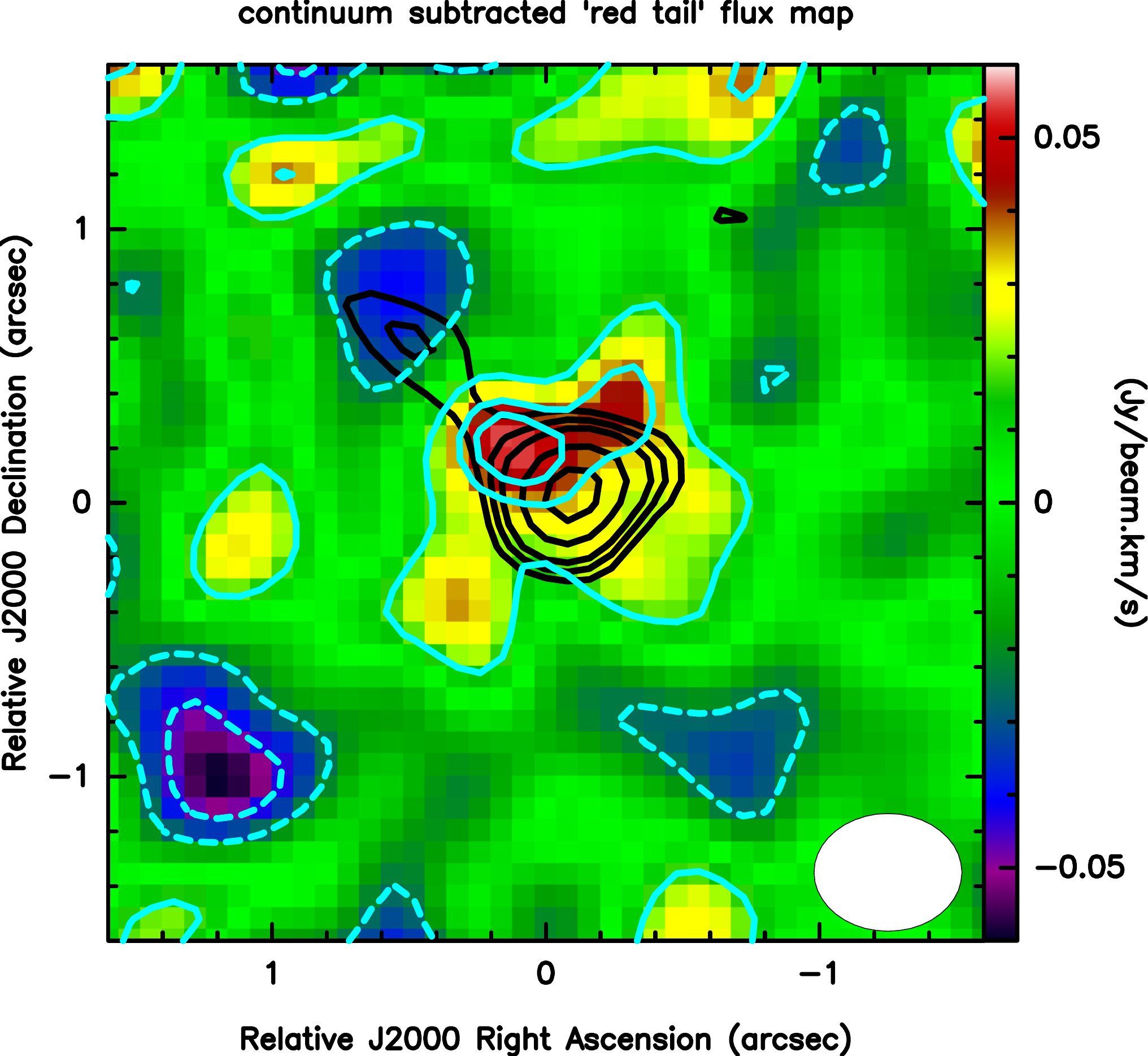}
 \caption{Flux maps extracted by collapsing the channels in the range [v$<-350$ km s$^{-1}$] (blue tail; left panel), and [v$>350$ km s$^{-1}$] (red tail; right panel). The images are extracted from the natural flux maps to maximise the sensitivity to detect faint features. The cyan contours represent the sigma levels: -1, -2, -3 (dashed) 1, 2, 2.5, and 3 (solid; 1$\sigma \sim $0.02 Jy km s$^{-1}$). The black contours at 3, 4, 5, 7, and $9\sigma $ indicate the dust continuum emission (from Fig.~\ref{continuum}). The beam ellipse is drawn in the lower right corner in both panels. The colour wedge gives the flux intensity scale in Jy km s$^{-1}$ beam$^{-1}$.}
         \label{COtails}
   \end{figure*}
%

In the left panel of 
Fig.~\ref{Xshooter} we compare the [O III]5007 line emission extracted from the central 1\arcsec aperture of the X-shooter data (red curve in the upper panel, arbitrarily normalised; \citealt{Perna2015}) with the CO(5-4) line profile (red histogram in the lower panel), taken from Fig. 2 and rebinned at 50 km s$^{-1}$.
For the [O III] line, three Gaussian components were needed to reproduce the line profile (black dashed curves in the upper left panel). The line profile of the CO(5-4) emission shows less conspicuous asymmetries compared with the [OIII] line. However, also in this case, we better model the total profile using three Gaussian components (black dashed curves in the lower left panel with FWHM of 190, 480 and 150 km s$^{-1}$, from the most blueshifted to the most redshifted). 

We also compared the total CO(5-4) profile with the line profile expected in the case of pure rotation that was obtained with $^{3D}$BAROLO (described in Sect.~4).   Given the uncertainties in the inclination angle, we report three model spectra extracted from the same aperture region as the CO(5-4) spectrum from the model data cubes reconstructed assuming 20, 30, and 40 deg inclination. The rotating disc models reproduce the total line profile reasonably well, although with residuals in the red section of the line. 

We further investigated the presence of high velocity molecular gas via the analysis of flux maps integrated at the blue and red tails of the CO line.
We constructed flux maps of the blue tail of the line ([v$<$ -350 km s$^{-1}$]; blue tail) and the red tail of the line ([v$>$ 350 km s$^{-1}$]; red tail)\footnote{We did not consider the channels in the range |300-350| km s$^{-1}$ in order to avoid overlapping channels in the images.}. For the detection
of faint features, we used the natural weighting scheme to construct these maps to maximise the sensitivity. The two panels of Fig.~\ref{COtails} show the continuum-subtracted flux maps extracted at |v|$> $350 km s$^{-1}$;  the dust continuum contours are superimposed as a reference for the nucleus and CO(5-4) line peak position. 
The  integrated maps of the  blue (peak emission at 3$\sigma$ level) and red (peak emission at 2.5$\sigma$ level) channels are not co-spatial with the dust continuum peak and are extended/elongated in opposite directions. The blue tail reaches scales of $\sim1.5$\arcsec\ from the nucleus ($\sim 13$ kpc). The red high velocity flux map  is instead located at a projected distance of $<0.5\arcsec$ ($<4 .5$ kpc) from the nucleus of the galaxy. 

These blue- and redshifted components cannot be due to rotational motions within the host galaxy, as the rotational major axis is roughly perpendicular to the position of the blue-red tails (see Fig.~\ref{vmap}). We note that 
the redshifted emission is  located between the nucleus and dust plume, and it could alternatively be associated with this continuum feature, tracing a minor merger. The excess emission at velocities larger than 300 km s$^{-1}$ observed in the host galaxy spectrum with respect to the rotation models (see left panel of Fig.~\ref{Xshooter})  may be associated with this component.
However, this minor merger  scenario would not explain the high velocity blueshifted gas observed in the opposite direction with respect to the nucleus.  

\begin{figure}
   \centering
 \includegraphics[width=8.7cm,angle=0]{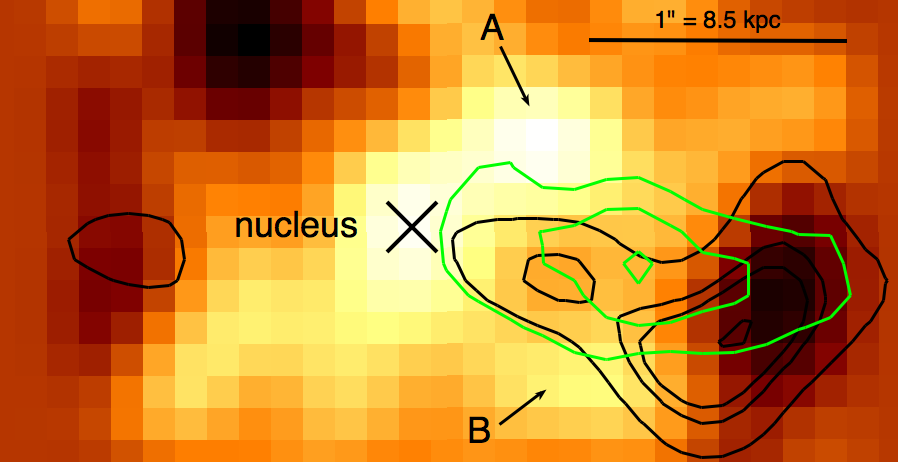}
      \caption{Narrow H$\alpha $ map (from Cresci et al. 2015a) with the contours of the molecular outflow (black; from the left panel of Fig.~\ref  {COtails}) and the contours of the ionised outflow  superimposed (green; from \citealt{Cresci2015}).  ALMA contours are drawn at 1, 2, 2.5, and 3$\sigma $. The cross indicates the QSO nucleus position. We also label with A and B the two star forming clumps reported in \citet{Cresci2015}. See text for details.}
         \label{halpha}
   \end{figure}

The flux maps shown in Fig. 9 closely resemble those expected in outflows events or detected in high S/N data (e.g. the Planck dusty Gem "Garnet", \citealt{Nesvadba2016}; the z=6.2 QSO J1148, \citealt{Cicone2015}). 
Despite the lower S/N of our data,  we note that the blue tail of the CO(5-4) is co-spatial with the observed [O III] blue emission ascribed to outflowing warm ionised gas. This is shown in Figure~\ref{halpha} in which the contours of the blue tail CO(5-4) emission (from the left panel of Fig.~\ref{COtails}; blue) are overplotted to  the narrow H$\alpha$ map obtained by SINFONI tracing young ($\lsimeq$10 Myr) SF regions in the galaxy and compared to the contours of the continuum-subtracted line map extracted from the blue wing of the [O III] emission (taken from \citealt{Cresci2015}; green). Although indirectly, this spatial coincidence reinforces the outflow interpretation as the origin of the observed high-v CO emission. If confirmed by deeper data, this would be the first time that, at high redshift, an outflow in both the molecular and ionised components has been resolved in the same object and on the same spatial region, therefore likely associated with the same outflow episode.

Finally, to assess the significance of the detection of high velocity molecular gas associated with the ionised outflow,  we extracted a spectrum
from a polygonal region encompassing the 1$\sigma$ contour of the blue outflow shown in Fig~\ref{COtails} and corresponding to the spatial region in which the [OIII] blueshifted putflow has been detected \citep{Cresci2015}. 
In the right panel of Fig.~\ref{Xshooter} we show  the CO(5-4) spectrum extracted from this region (red histogram),  
the $\pm$1$\sigma$ level  (dotted lines), and the average of three noise spectra taken randomly in the field over a region of the same area and shape of that of the outflow (purple circles). 
Although the emission around zero velocity is dominated by the disc component, we detect emission clearly above the noise in channels not dominated by the rotation (from $\sim$500 to $\sim$800 km s$^{-1}$, both in the blue and red section); we overplot the spectrum
taken from
the left panel for comparison as a black open histogram.   Emission at these velocities is detected only in the outflow spectrum and not in the nuclear spectrum.  
The significance of the detection of the feature in the blue (red) channels is S/N=4.55$\sigma$ (2.50$\sigma$), assuming a Gaussian function with FWHM=280 (260) km s$^{-1}$ (blue curves in Fig.~\ref{Xshooter}). The maximum  blueshifted (redshifted) velocity observed is of the order of v$\sim-700$ ($+$900) km s$^{-1}$.  We checked that the significance of the detection does not change if we extract the outflow spectrum avoiding the central beam. 
The total flux associated with the two components is I$_{\rm CO(5-4),blue+red}=0.11\pm0.023$ Jy km s$^{-1}$.

Under the asssumption that the high velocity CO(5-4) emission is tracing outflowing gas, we assume the same gas excitation ratio and $\alpha_{\rm CO}$ values discussed in Section 5 to derive the outflowing mass from the flux associated with the outflow. This translates into a total gas mass associated with the outflow, M$_{\rm gas,out}\sim1.4\times10^9$ M$_\odot$. 
If we instead assume $\alpha_{\rm CO}$ is 0.13, as suggested by recent numerical simulations with molecules  formed and accelerated in situ in AGN-driven galactic winds \citep{Richings2018}, we derive an outflowing mass M$_{\rm gas,out}\sim0.2\times10^9$ M$_\odot$.  We stress that the two values suffer from large uncertainties, given the chain of assumptions employed, and should therefore be considered as a representative estimate of the order of magnitude of the outflowing molecular gas mass. 
Assuming a spatial scale of 10 kpc and an outflow velocity of 700 km s$^{-1}$ as inferred from our data, and in the hypothesis of uniform density in the emitting region (see \citealt{Cresci2015,Fiore2017}), we estimate an average mass outflow rate $\dot{M}_{gas,out}\sim 50-350$  M$_\odot$ yr$^{-1}$  in the case of $\alpha_{\rm CO}$=0.13 and of $\alpha_{\rm CO}$=0.8, respectively. 
In Cresci et al. (2015a) the mass outflow rate of the ionised gas component has been reported to be $>300$ M$_\odot$ yr$^{-1}$. In \citet{Perna2015} it was also reported a lower limit of the neutral  gas outflow of $>80$  M$_\odot$ yr$^{-1}$ based on MgII absorption line detected in X-shooter data. 
Summing up all the gas components, the total mass rate of the outflowing gas would therefore be $\dot{M}_{out, tot}\sim500-800$ M$_\odot$ yr$^{-1}$.  We note that this value is lower than that reported in Cresci et al. (2015a; $>1000$ M$_\odot$ yr$^{-1}$), where a correction factor $>3$ to infer the total outflowing gas mass from the ionised gas measurement has been applied, based on results of outflowing gas in local galaxies (see e.g. \citealt{Carniani2015}). 
Our estimate of the molecular gas outflow component points towards a comparable contribution to the mass outflow rate for the gas in this phase. This appears in line with the findings recently presented by \citet{Fiore2017}, who pointed out that at high AGN bolometric luminosities (L$_{\rm bol}>10^{46}$ erg s$^{-1}$) outflow rates measured in the ionised and molecular gas components could be comparable. 

We finally note that if we take into account the total gas consumption due to the outflow, the depletion timescale would further reduce, t$_{\rm depl}$=M$_{\rm gas}$/(SFR+ $\dot{M}_{\rm out, tot}$)$\sim13-30$ Myr. We plot the new estimate of the depletion timescale inferred from the CO line luminosity as pink point in Fig. 7, which further enhances the deviation of XID2028 from the locus expected from host galaxies of the same sSFR properties because of a combination of active SFR and effective gas ejection from the central region of the galaxy.

%

\section{Summary and conclusions}

We presented ALMA data of  XID2028, a  star forming-QSO system at high redshift (z$\sim1.5$), thought to be in the feedback phase. The presence of a massive outflow with significant impact on the host galaxy has been previously unveiled through NIR observations \citep{Perna2015,Cresci2015}, and anomalous molecular gas content has been reported from the study of the CO(3-2) line transition \citep{B15_PDBI}. 

With the new analysis at (sub)millimetre wavelengths, also informed by high-resolution near-infrared imaging obtained through LUCI@LBT and complete multiwavelength coverage in the framework of the COSMOS survey, we report the following results:
\begin{itemize}
\item[$\bullet$] We detected compact ($\sim$2.5 kpc diameter) dust continuum emission at rest-frame $\sim500 \mu$m, confined to a central region of the galaxy and coincident with a point source detected by HST (at $\sim0.1\arcsec$ resolution) and LBT/LUCI (at $\sim0.27\arcsec$ resolution; see Sections 2.1 and 3; Fig.~\ref{continuum} and \ref{hst}). The continuum fluxes in ALMA bands 6 and 7, along with the far-infrared  data points from Herschel, constrain the dust temperature at T$_{\rm dust}\sim50$ K (see Section 5 and Fig.~\ref{SED}). \\
 
\item[$\bullet$] We detected the CO(5-4) transition at $\sim30\sigma$ significance (see Section 2.2 and Fig. 2). Also in this case, the emission is confined to the central $\sim$2.5 kpc diameter area. The CO(5-4), dust continuum, rest-frame optical, and AGN nuclear emissions are all coincident with and originate from the same central region within the limits of our observations. \\

\item[$\bullet$] We detected an extended emission in the dust continuum  elongated towards the NE up to $\sim10$ kpc from the nucleus, coincident with an asymmetry seen in the K-band continuum from LUCI data (see Fig.~1 and Sect. 2.1 and 3). This component could be in principle ascribed to a minor merger or a tidal tail. \\

\item[$\bullet$] The bulk of the molecular gas associated with the CO(5-4) emission appears to be structured into a rotating molecular disc (see Section 4 and Fig. 5,6) of the order of few kpc in radius (more compact than the size measured in the rest-frame optical light, $>5$ kpc), and with an inferred  rotation velocity of the order of $\sim400$ km s$^{-1}$. \\

\item[$\bullet$] 
We have been able to substantially refine the measurement of the gas mass
(M$_{\rm gas}=1.1\pm0.5\times10^{10}$ M$_\odot$), gas fraction ($<5$\%), and
 depletion timescale of converting gas into stars (SFR/M$_{\rm gas}$,  t$_{\rm
depl}$=40-75 Myr) thanks to a more robust analysis enabled by the combination of the ALMA data, CO SLED extrapolations, and spatial resolution. This measurement
firmly places these values towards the lower range of the previous estimates based only on the PdBI data reported in Brusa et al. (2015b) and thereby reducing the uncertainties of a factor of 2 to 5 (see Sections 2.4 and 5, and Figs.~\ref{COSLED}, 7). A comparison of the depletion timescale of XID2028 with other AGN with outflows and normal SF properties seems to reinforce the idea that anomalous gas conditions and consumptions are related with the presence of outflow episodes and/or AGN activity, possibly a consequence of feedback from the BH.  \\

\item[$\bullet$]  We report a $\sim4.5\sigma$ detection of high velocity CO emission with velocities approaching as high as -700 km s$^{-1}$ (see Fig.~\ref{Xshooter} and \ref{COtails}; see Section 6). 
The high velocity components of the CO(5-4) line emitting regions cannot be due to rotational motion within the host galaxy, as they are  detected in regions perpendicular to the rotation axis of the galaxy.  We also report evidence (albeit at a lower significance level of  $\sim2-3\sigma$) of spatially resolved emission for these high velocity CO(5-4) components. The emission in the high velocity channels (including the red tail) can be interpreted as outflowing gas  in opposite directions, as predicted in biconical outflow models. \\

\item[$\bullet$] We report that the blueshifted emission appears to be co-spatial with the ionised outflow seen in SINFONI data (see Fig.~\ref{halpha}).  XID2028 may therefore represent a unique case beyond the local Universe with a spatially resolved outflow in the molecular and ionised phase, over the same spatial region, and  therefore likely associated with the same outflow episode (different from, e.g. SDSSJ1356 reported in \citealt{Sun2014}).  With this assumption, we report a total mass outflow rate of $\dot{M}_{\rm tot}\sim500-800$ M$_\odot$ yr$^{-1}$.  To our knowledge, this is the first time that the measurement of total outflowing mass including 
 the molecular and atomic components, in both the ionised and neutral phases, is attempted for a z$\sim1.5$ QSO.

\end{itemize}

All the physical properties of XID2028 discussed in this work are summarised in Table 2.

XID2028 shows depletion timescale similar to those of bright submm galaxies, which are in contrast forming stars at much higher rates ($>1000-2000$ M$_\odot$ yr$^{-1}$ versus $\sim270$ M$_\odot$ yr$^{-1}$ measured for XID2028) and are characterised by a much larger gas reservoir ($>20-50$\% gas fraction versus $<$5\% measured for XID2028).  The fact that obscured SFR in XID2028 is still detected, as witnessed by the strong FIR emission and the reddened SED,  fits an evolutionary scenario in which XID2028 is observed in a later stage of the merger sequence with respect to SMGs, when the SFR is considerably dropped down, but is still ongoing in the central regions where the BH is also actively fed. 
The residual gas reservoir will eventually be exhausted and it is probable that in  $\sim10$ Myrs  XID2028 will turn into a massive elliptical system with no more fresh fuel to form stars.

In terms of FWHM size and high rotation velocities inferred for the molecular disc component, XID2028 resembles other high-redshift compact star forming galaxies (cSFGs; \citealt{Barro2014}) recently reported in the literature on the basis of high-resolution ALMA observations \citep{Tadaki2017}). Similar to what has been proposed for GMASS-953 \citep{Popping2017}, the nuclear starburst may have brought the gas to the centre, fuelling both SF and nuclear activity. Simulations predict that, before it is completely depleted (via efficient SF and/or QSO winds), this gas should be rotating at high velocity while setting into the nuclear region (see e.g. \citealt{Shi2017}). The high rotational velocity observed in XID2028 ($\sim400$ km s$^{-1}$) along with the short depletion timescale and the presence of a powerful outflow episode seem to nicely fit this scenario. All the observational results therefore point towards the fact that the outflowing wind may be the result of the concurrence of a powerful AGN and a  starburst episode in a compact nuclear region. 

A systematic study aimed at linking in a quantitative way  AGN, hosts, and outflow properties for objects in which the outflow component has been already revealed is the only way to improve our understanding of the QSO outflow phenomenon and its consequence on the galaxy growth. The lack of dedicated observations of such rare sources may be the reason why, in addition to sensitivity limitations, neutral and molecular outflows at high-z have been reported only in the more intense [CII] line and only in few z$>$ 2 QSOs (e.g. \citealt{Wagg2012,Maiolino2012}, see section 4.7 in \citealt{CW2013}), while they are now seen routinely in observations of the ionised gas component. 

As a final note, we caution that recent results from CO blind surveys such as ASPECS \citep{Walter2016,Decarli2016} suggest that the picture we have drawn for CO excitation ratios from dedicated CO observations of known unobscured QSOs may not be representative of the entire AGN population. For example, the three X-ray detected AGN in  Decarli et al. (2016; ID 1, ID 2, and ID 7, all with X-ray luminosities between 5$\times10^{42}-5\times10^{43}$ erg s$^{-1}$) have very different CO SLED. Of the three, ID7 is the closer to XID2028 in terms of host galaxy properties (SFR, t$_{\rm depl}$, $\Sigma_{\rm SFR}$,M$_{\rm H2}$; see Table 3 in Decarli et al. 2016), although it has a factor of $\sim30$ lower AGN bolometric luminosity, and still it has a CO SLED that is virtually consistent with the MW SLED. 
Therefore, CO SLEDs studies on large AGN samples, especially out to high J-transitions, will be crucial to assess the role of AGN ionisation in shaping the properties of the molecular gas reservoirs of AGN host galaxies.

\begin{acknowledgements}
This paper makes use of the following ALMA data: ADS/JAO.ALMA\#2015.1.00299.S (PI: Brusa), \#2015.1.00137.S (PI: Scoville), and \#20015.1.00171.S (PI: Daddi). ALMA is a partnership of ESO (representing its member states), NSF (USA) and NINS (Japan), together with NRC (Canada), NSC, and ASIAA (Taiwan), and KASI (Republic of Korea), in cooperation with the Republic of Chile. The Joint ALMA Observatory is operated by ESO, AUI/NRAO, and NAOJ. XID2028 is an interesting target from several perspectives (e.g. massive MS galaxy; bright FIR/Herschel source with substantial SFR), and as such, in addition to our  deep dedicated follow-up,  it has also been targeted independently by two different ALMA snapshot programmes as part of large samples  (100+) of high-z galaxies.  A careful scrutiny of the ALMA archive and an open collaboration with the PIs of the programmes allowed us to make the best use of  ALMA data.
We thank the ARGOS team for obtaining the K-band imaging of XID2028 during a commissioning run. We gratefully thank the staff of the Italian ARC node for their support in data reduction. 
This research was finalised at the Munich Institute for Astro- and Particle Physics (MIAPP) of the DFG cluster of excellence ``Origin and Structure of the Universe", during the programme ``In \& Out: What rules the galaxy baryon cycle?" (July 2017). 
MB, MP, and GL acknowledge support from the FP7 Career Integration Grant ``eEASy'' (``SMBH evolution through cosmic time: from current surveys to eROSITA-Euclid AGN Synergies'', CIG 321913). EL is supported by a European Union COFUND/Durham Junior Research Fellowship (under EU grant agreement no. 609412). MTS was supported by a Royal Society Leverhulme Trust Senior Research Fellowship (LT150041). CF acknowledges support from  the European Union Horizon 2020 research and innovation programme under the Marie Sklodowska-Curie grant agreement No 664931. SC  acknowledges financial support from the Science and Technology Facilities Council (STFC). We acknowledge financial support from INAF under the contracts PRIN-INAF-2014 (``Windy Black Holes combing Galaxy evolution''). We gratefully thank Maurilio Pannella, Renzo Sancisi, Margherita Talia, Anna Cibinel, Roberto Decarli,  Andrea Lapi, Gerg\"o Popping, David Rosario, Kartik Sheth, and John Silverman for useful discussions, and Darshan Kakkad for providing data prior to publication. We thank the anonymous referee for her/his constructive criticisms, which considerably improved the presentation of the results.

\end{acknowledgements}

\appendix 
\section{Spectral energy distribution of XID2028}

\begin{figure}
   \centering
 \includegraphics[width=8.7cm,angle=0]{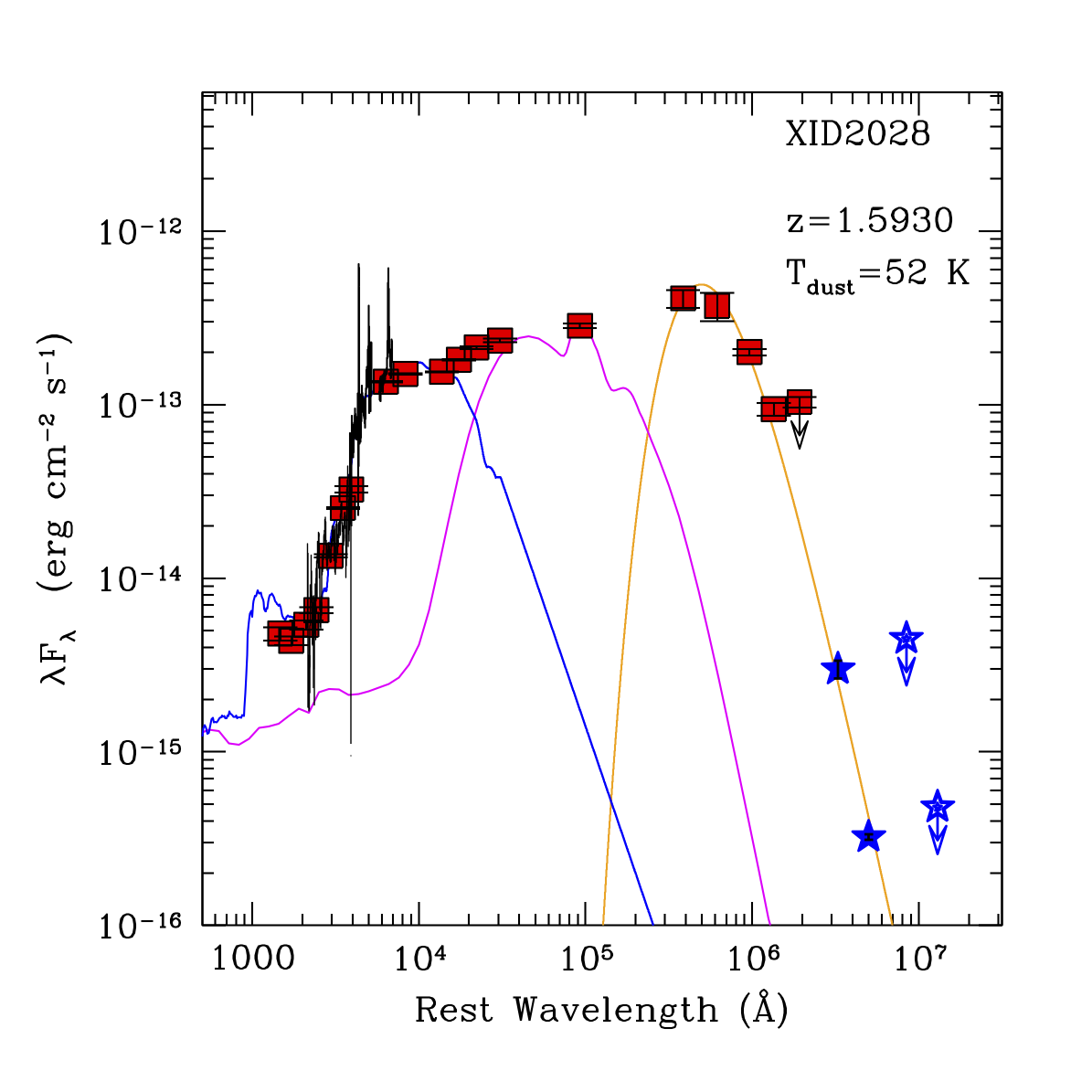}
      \caption{Spectral energy distribution of XID2028, shown in the rest frame. Red squares indicate the observed photometry, in the UV-optical-NIR-MIR-FIR regime, extracted from the Laigle et al. (2016) catalogue. The blue stars indicate the ALMA bands 7 and 6 data points and 2mm and 3mm upper limits from PdBI and ALMA Band 3. In the rest-frame optical range we also over plot the X-shooter spectrum (from \citealt{Perna2015}). The blue, magenta, and orange curves correspond to the stellar, AGN, and grey-body components found as a best-fit solution of the overall photometry. }
         \label{SED}
   \end{figure}

The result of the SED fit to the COSMOS (Laigle et al. 2016), PdBI, and ALMA photometry, performed to derive the dust temperature, is shown in Fig.~\ref{SED}. 
We note that from this SED fit we derive a host galaxy stellar mass M$_\star\sim10^{12}$ M$_\odot$ and a star formation rate SFR=240 M$_\odot$ yr$^{-1}$. We also ran the publicly available fully Bayesian Markov Chain Monte Carlo SED fitting algorithm {\it AGNfitter} \citep{Agnfitter}.
We considered the fiducial template library supplied with {\it AGNfitter}, from which we included the additional ALMA bands to better model the Rayleigh-Jeans part of the SED. 
With {\it AGNfitter} we  obtained results very consistent with these values of host galaxies parameters (M$_\star\sim10^{12}$ M$_\odot$  and SFR=280 M$_\odot$ yr$^{-1}$). In both cases, the emission observed in the NIR range is explained solely by the host galaxy component (blue curve in Fig. ~\ref{SED}). Similarly, \citet{Laigle2016}, who fit the observed photometry without an AGN contribution, provide a stellar mass of M$_\star$=$1\pm0.1\times10^{12}$ M$_\odot$.

These new values of the host galaxy parameters should be compared with those presented in \citet{Perna2015} and adopted in \citet{B15_PDBI} (reported in Table 2).
The SFR measurement is stable in the various fits (see also \citealt{Bongiorno2014}), while the stellar mass is a factor of 2 higher in the new fits. This highlights the uncertainty associated with the degeneracy of various components in SED fitting routines. In this work we consider as fiducial stellar mass value M$_\star$=$4.5\times10^{11}$ M$_\odot$, also motivated by the fact that our LUCI imaging suggests a resolved host galaxy contribution to the K-band flux $>20$\% (see Section 3). In the unlikely case that the remaining 80\% unresolved flux is entirely due to the central AGN, we conservatively assume as a lower limit of the stellar mass M$_\star$= 2$\times$10$^{11}$ M$_\odot$. We take into account of all uncertainties affecting SFR (SFR=240-280 M$_\odot$ yr$^{-1}$) and M$_\star$ (from 2$\times$10$^{11}$ to 10$^{12}$ M$_\odot$) in the error associated with the sSFR. We however note that the value of T$_{\rm dust}$ is constrained by the Herschel and ALMA data points and does not depend on  the details of the fit at shorter wavelengths.

We stress that the results presented in the paper would not change significantly assuming a lower stellar mass. In particular, the gas fraction would remain $<10$\% if we consider the lower limit of our estimate of the stellar mass (2$\times10^{11}$ M$_\odot$). Assuming a higher stellar mass ($10^{12}$ M$_\odot$) instead would move XID2028 further to the left in Fig. 7. 

Finally, \citet{Symeonidis2016} pointed out that the intrinsic AGN emission could contribute significantly to the FIR emission and therefore the SFR quoted above is likely to be considered an upper limit. However, given the SED shape between 5 $\mu$m and 100 $\mu$m we expect that the SFR for XID2028 derived when assuming the \citet{Symeonidis2016} template would be at most a factor of 2 lower.  A change of a factor 2 in the SFR would only slightly move XID2028 towards the up-left direction in Fig. 7, but its deviation from the scaling relations expected for normal star forming galaxies would still be a factor of $\sim10$. 

%

\end{document}